\documentclass[10pt]{iopart}

\usepackage{cite,graphicx}

\newcommand{\twid}{\sim}

\newcommand{\paren}[1]{\left( #1 \right)}
\newcommand{\square}[1]{\left[ #1 \right]}

\newcommand{\casesshortii}[4]
{\left\{
\begin{array}{ll}
#1\ \ ,\gap & #2 \\  #3\ \ ,\gap & #4 
\end{array}%
\right.
}

\newcommand{\casesbracketsshortii}[4]
{\left\{
\begin{array}{ll}
#1\ & (#2) \\  #3\ & (#4) 
\end{array}%
\right.
}

\newcommand{\gap}{\hspace{.4in}}

\newcommand{\gt}{\rightarrow}

\newcommand{\period}{\ \ .}
\newcommand{\comma}{\ ,\ }

\newcommand{\gsim}{\,\stackrel{>}{\scriptstyle \sim}\,}

\newcommand{\ignore}[1]{}

{\end{list}}

{\end{list}}

{\ \\ XXXXX \hfill XXXXX #1 XXXXX\begin{equation}\label{#1}}%
{\end{equation}}

{\ \\ XXXXX \hfill XXXXX #1 XXXXX\begin{eqnarray}\label{#1}}%
{\end{eqnarray}}

\newenvironment{bequation}[1]%
{\begin{equation}\label{#1}}%
{\end{equation}}

\newenvironment{beqnarray}[1]%
{\begin{eqnarray}\label{#1}}%
{\end{eqnarray}}

\newcommand{\drop}{\nonumber \\}

\newcommand{\ie}{i.\,e.~}
\newcommand{\eg}{e.\,g.\@ }

{\begin{center} \section*{#1} \end{center}}

{\end{figure}} 

{\end{figure}}

{\end{figure}}

 \def\eqref#1{(\ref{#1})}

\newenvironment{eq}[1]%
{\begin{bequation}{#1}}{\end{bequation}}

\newenvironment{eqarray}[1]%
{\begin{beqnarray}{#1}}{\end{beqnarray}}

\renewcommand{\textheight}{700pt}

\def\eref#1{equation (\ref{#1})}

\newcommand{\taus}{\tau_s}

\newcommand{\phistar}{\phi^{*}}

\newcommand{\Gammabound}{\Gamma_{\rm bound}}

\newcommand{\taubulk}{\tau_{\rm bulk}} 
 
\newcommand{\ts}{t_{\rm s}} 
\newcommand{\tauex}{\tau_{\rm ex}} 
\newcommand{\taueq}{\tau_{\rm eq}} 

\newcommand{\RF}{R_F}

\newcommand{\ta}{t_a}

\newcommand{\ncont}{n_{\rm cont}}

\newcommand{\rhosuper}{\rho_{\rm super}}
\newcommand{\Gammaboundinf}{\Gamma_{\rm bound}^{\infty}}
\newcommand{\Gammainf}{\Gamma^{\infty}}

\newcommand{\lsep}{l_{\rm sep}}

\newcommand{\tauads}{\tau_{\rm ads}}

\newcommand{\nth}{^{\rm th}}

\newcommand{\sstar}{s^*}

\newcommand{\Zbar}{\bar{Z}}
\newcommand{\Xibar}{\bar{\Xi}}
\newcommand{\sbarl}{\bar{s}_l}
\newcommand{\sbart}{\bar{s}_t}
\newcommand{\zstar}{z^*}
\newcommand{\tfinal}{t_{\rm final}}
\newcommand{\hprox}{h_{\rm prox}}
\newcommand{\Peq}{P^{\rm eq}}
\newcommand{\fbar}{\bar{f}}
\newcommand{\fmin}{f_{\rm min}}
\newcommand{\fmax}{f_{\rm max}}
\newcommand{\hexp}{h_{\rm exp}}
\newcommand{\epsilonc}{\epsilon_{\rm c}}
\newcommand{\Tg}{T_{\rm g}}
\newcommand{\Tm}{T_{\rm m}}

\newcommand{\Zl}{Z_l}
\newcommand{\Zt}{Z_t}
\newcommand{\Wl}{W_l}
\newcommand{\Wt}{W_t}
\newcommand{\Al}{A_l}
\newcommand{\At}{A_t}

\begin{document}



\topical{Non-Equilibrium in Adsorbed Polymer Layers}

\author{Ben O'Shaughnessy and Dimitrios Vavylonis}

\address{Department of Chemical Engineering, Columbia University, \\ New  York, NY 10027, USA}

\ead{\mailto{bo8@columbia.edu}, \mailto{dv35@columbia.edu}}

\begin{abstract}

High molecular weight polymer solutions have a powerful tendency to
deposit adsorbed layers when exposed to even mildly attractive
surfaces.  The equilibrium properties of these dense interfacial
layers have been extensively studied theoretically.  A large body of
experimental evidence, however, indicates that non-equilibrium effects
are dominant whenever monomer-surface sticking energies are somewhat
larger than $kT$, a common case.  Polymer relaxation kinetics within
the layer are then severely retarded, leading to non-equilibrium
layers whose structure and dynamics depend on adsorption kinetics and
layer ageing. Here we review experimental and theoretical work
exploring these non-equilibrium effects, with emphasis on recent
developments. The discussion addresses the structure and dynamics in
non-equilibrium polymer layers adsorbed from dilute polymer solutions
and from polymer melts and more concentrated solutions.  Two distinct
classes of behaviour arise, depending on whether physisorption or
chemisorption is involved.  A given adsorbed chain belonging to the
layer has a certain fraction of its monomers bound to the surface,
$f$, and the remainder belonging to loops making bulk excursions.  A
natural classification scheme for layers adsorbed from solution is the
distribution of single chain $f$ values, $P(f)$, which may hold the
key to quantifying the degree of irreversibility in adsorbed polymer
layers.  Here we calculate $P(f)$ for equilibrium layers; we find its
form is very different to the theoretical $P(f)$ for non-equilibrium
layers which are predicted to have infinitely many statistical classes
of chain.  Experimental measurements of $P(f)$ are compared to these
theoretical predictions.

\end{abstract}


\submitto{\JPCM}

\maketitle


\section{Introduction}

High molecular weight polymers are extremely surface-active molecules.
Even a weak interaction between a single monomer and a surface tends
to be magnified into a powerful attraction or repulsion when many of
these monomers are linked together to form a polymeric chain.  It is a
remarkable fact that surfaces contacting even extremely dilute polymer
solutions can readily develop very dense polymer layers.  Technologies
revolving around the properties of either synthetic or biological
polymer layers are many and varied, including adhesion
\cite{wool:poliface_book,wu:polymer_iface_adhesion_book}, coating
\cite{wicks:book_coatings_2nd}, colloid stabilization \cite{russel:colldisp}, fiber
reinforced thermoplastics
\cite{mittal:book_adhesion,edwards:review_filler_reinforcement},
flocculation processes \cite{hogg:flocculation_polads_99}, DNA
microaarrays \cite{brownbotstein:cdna_array} and biocompatibilization
\cite{malmsten:biopol_ifaces_collection}.  Motivated both by
fundamental questions and by technology, understanding and predicting
the structure and formation kinetics of these soft layers is a major
concern of polymer science
\cite{granick:macromolecules_surfaces_review_03}.  A crucial aspect of
experimental studies is that adsorbed polymer dynamics are typically
extremely slow for long chains: an individual chain has many surface
stickers and interacts with numerous other chains impeding its motion.
Irreversibility and non-equilibrium effects are therefore very common.
The subject of this review is experimental and theoretical work on
these non-equilibrium effects, and though we consider adsorption from
dilute solutions, semi-dilute solutions and melts our main emphasis is
neutral homopolymer adsorption from dilute solutions. This is the
simplest and most widely studied class.

\begin{figure}[htb]
\centering
\resizebox{0.8\columnwidth}{!}{%
  \includegraphics{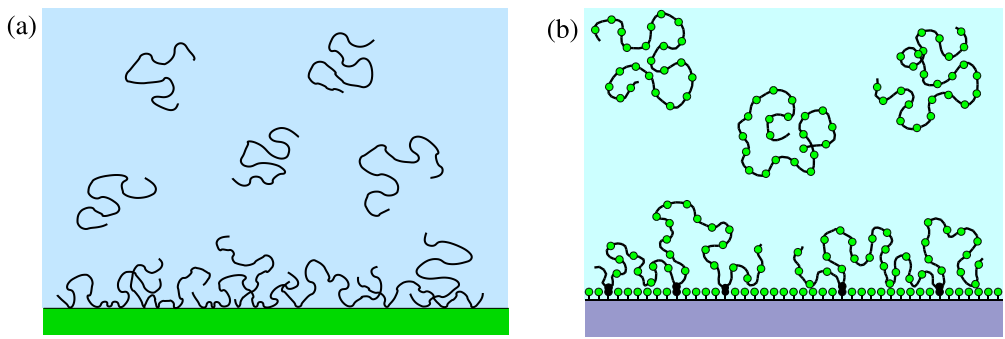}
}
\caption{(a) Schematic of physisorption from a fluid polymer phase
onto a surface.  Adsorbed chains consist of loops, tails and sequences
of bound monomers (``trains'').  When non-equilibrium effects become
important, layer structure depends on kinetics of adsorption.  This
review addresses phyisorption from dilute solutions in sections 2, 3
and 4 and physisorption from melts in section 6. (b) As in (a) but for
chemisorption.  In this case chains carry reactive groups which can
form covalent bonds (shown black) with a functionalized surface.
Chemisorption from dilute solutions is reviewed in section 5 and from
concentrated solutions in section 6.}
\label{scheme}      
\end{figure}

Polymer-surface adsorption systems are naturally classified according
to the mode of adsorption. Roughly, there are two classes:
chemisorption and physisorption (see \fref{scheme}).  The clearest
example of irreversibility arises in {\em chemisorption}
(\fref{scheme}(b)) where the covalent polymer-surface bonds which
develop are often essentially irreversible on experimental timescales.
Monomer sticking free energies, $\epsilon$, have values typical of
covalent bonds which are one to two orders of magnitude greater than
$kT$.  Chemical adsorption is employed in various technologies where
polymers are attached by chemical reactions to solid surfaces either
from a polymer melt as in the reinforcement of polymer-polymer or
polymer-solid interfaces
\cite{leger:anchored_chains_adhesion_review,wu:polymer_iface_adhesion_book,edwards:review_filler_reinforcement,creton:kramer:iface_fracture_review},
or from solution as in colloid stabilization by chemically grafting
polymers onto particle surfaces
\cite{laiblehamann:chemi_colloid,halperin:tethered_microstruct_rev,tirelllevivky:end_tethered_opinion_1997}.

What is less obvious is why non-equilibrium effects are so widely
observed in {\em physisorbing} systems, even for rather weak sticking
energies.  Available experimental evidence suggests that
irreversibility effects become important as soon as $\epsilon$ becomes
somewhat larger than $kT$.  For example the experiments by Schneider
et al. \cite{schneider:granick_bimodal,douglas:kinetics_pol_ads_jpc}
for polymethylmethacrylate (PMMA) adsorption onto oxidized silica via
hydrogen bonding in dilute CCl$_4$ solutions ($\epsilon \approx 4 kT$)
show essentially frozen-in adsorbed chain configurations.  Large
physisorption sticking energies ($\epsilon > kT$) originate in
hydrogen bonding or other dipolar forces, dispersion forces or
attractions between charged groups. Metal and silicon-based surfaces
are usually oxidized and many polymer species form strong hydrogen
bonds with the surface oxygen or silanol groups
\cite{vanderbeek:segmental_polads_energies_91,frantzgranick:silicon_surface_preparation}.
Biopolymers such as proteins and DNA attach tenaciously to many
surfaces due to their many charged, polar and hydrophobic groups
\cite{malmsten:biopol_ifaces_collection,brownbotstein:cdna_array,hladybuijs:protein_adsorption}.
Since hydrogen bonds, for instance, typically have energies of several
$kT$ \cite{israelachvili:book,joestenschaad:book:h_bonding} it is
apparent that strong physical bonds are very common.  This suggests
that whether physical or chemical bonding is involved, for long chains
irreversible effects may in practice be the rule rather than the
exception.

To understand non-equilibrium layers, one must identify how they
differ from equilibrium layers.  The theory of fully equilibrated
layers is far more advanced, at both the mean-field
\cite{semenov:loops_tails_macro} and scaling
\cite{gennes:self_similar_profile,gennes:pols_interface_review,
eisenriegler:proximal_exponent,eisenriegler:pol_adsorption_scaling_monte_carlo,
gennespincus:proximal_exponent,gennes:ads_pol_loops,aubouy:scaling_flat_layers,semenovjoanny:loops_tails_europhys}
level of description.  A main result of these theories is expressions
for the decay of the monomer density profile as a function of the
distance $z$ from the surface.  For adsorption from dilute solutions
for example, in the scaling picture originally developed by de Gennes
\cite{gennes:self_similar_profile,gennes:pols_interface_review},
Eisenriegler et al.
\cite{eisenriegler:proximal_exponent,eisenriegler:pol_adsorption_scaling_monte_carlo},
and de Gennes and Pincus \cite{gennespincus:proximal_exponent}, each
adsorbed chain has sequences of surface-bound monomers (trains)
interspersed with portions extending away from the surface (tails and
loops of size $s$) with distribution $\Omega(s)\twid s^{-11/5}$
\cite{gennes:ads_pol_loops,aubouy:scaling_flat_layers,semenovjoanny:loops_tails_europhys}
leading to a self-similar density profile $c(z) \twid z^{-4/3}$.
Experimentally, the existence of an extended diffuse layer is well
established by a large number of neutron scattering 
\cite{cosgrove:neutron_scattering_ads_pol_1987_macro,cosgrove:neut_refl_block_homo_mica,
cosgrove:pol_ads_profiles_review_1990,auvraycotton:neutron_scattering_adsorbed_pol,
rennie:neut_refl_peo_air_water,hone:cosgrove_neut_scat_peo} and
neutron reflectivity \cite{lee:neutron_reflectivity_adsorbed_pol,guiselin:neut_refl_ads_pol_jcp,
sun:pol_profile_neutron_reflection} studies.  However a universally accepted
quantitative test of the predicted density profiles has been difficult
to achieve, both due to intrinsic limitations of the experimental
techniques \cite{fleer:pol_iface_book} and to the asymptotic nature of
many of the theoretical results which are valid in the limit of very
long chains.  Furthermore, for experiments specifically directed at
equilibrium, ensuring that equilibrium conditions are realised is
difficult when the very non-equilibrium effects one wishes to avoid
are poorly identified.

Understanding the origin of the observed deviations from equilibrium
for weakly adsorbing systems in dilute solutions is a major unresolved
puzzle in polymer science.  At present it is unclear how universal
these non-equilibrium effects may be.  Various effects have been
speculated to play important roles.  Kremer
\cite{kremer:ads_pol_glassy} and de Gennes
\cite{gennes:ads_pol_dynamics_toyota} have suggested that if the
solution temperature is below the polymer melt glass transition
temperature, glassy effects may onset at the interface where polymer
densities are similar to those of melts.  Interactions with the
surface might further enhance glassiness.  Chakraborty and coworkers
\cite{chakrabort:glassy_polymer_solid,
chakrabortyadriani:glassy_polymer_solid_2,
shafferchakraborty:pmma_chemisorption_kinetics,
adrianichakraborty:kinetic_ising_chemisorption} suggested
flattened-down chains experience strong kinetic barriers due to local
steric constraints which drastically slow down dynamics.  Ideas
related to slow cooperative motions, mutual pinning, development of
entanglements at the interface and crystalization have been proposed
by Johner and Semenov \cite{johnersemenov:comment_polymer_mat}, Sommer
\cite{sommer:comment_polymer_mat}, Granick
\cite{granick:comment_polymer_mat} and Raviv et el.
\cite{raviv:reply_comments_polymer_mat} in a series of comments
following a recent theoretical work by Raviv et al.
\cite{raviv:polymer_mat} which interpreted past experiments
\cite{luckhamklein:surface_force_peo_ps,raviv:surface_force_shear}
exhibiting non-equilibrium effects.

In this review we do not attempt an exhaustive review of the vast body
of past research work involving strongly physisorbing or chemisorbing
polymers at interfaces.  Instead, with fundamental issues in mind, our
aim is to (i) assemble and classify a number of theoretical ideas and
numerical simulations which articulate the community's current level
of understanding of equilibrium and non-equilibrium polymer
adsorption, and (ii) summarize a number of experimental results which
we believe are particularly pertinent and which still demand a
theoretical explanation.  The emphasis is on the simplest case:
adsorption of neutral flexible polymers from dilute bulk solutions.
We also review work on irreversible adsorption from semi-dilute bulk
solutions and melts, motivated by ideas initiated by Guiselin
\cite{guiselin:irrev_ads}.  Polyelectrolyte solutions, polymers with
complex architectures and non-flat surfaces are beyond the scope of
the present review.

\begin{figure}[tb]
\centering
\resizebox{0.45\columnwidth}{!}{%
  \includegraphics{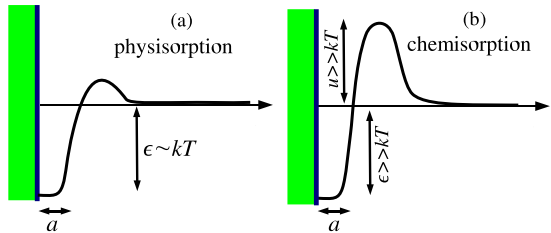}
}
\caption{The two broad classes of polymer adsorption, physisorption
and chemisorption, have very different values of the parameter $Q$,
the local monomer-surface association rate. $Q$ can be thought of as
the conditional monomer-surface sticking probablity per unit time,
given the unattached monomer contacts the surface.  Though many
systems are in practice mixtures of chemisorption and physisorption, a
simplified view of monomer free energy as a function of distance
between monomer and surface is shown.  (a) For physisorbing polymers,
the activation barrier is very small and and monomer-surface
association is very likely upon contact, \ie $Q \ta$ is of order
unity, where $\ta$ the monomer relaxation time.  When the sticking
energy $\epsilon$ exceeds a few $kT$, experiment indicates that chains
need large amounts of time to escape the surface, presumably due to
complex many-chain effects.  (b) Chemisorption typically involves a
large activation barrier, $u \gg kT$.  Many monomer-surface collisions
are needed to traverse this barrier, $Q \ta \ll 1$.  The adsorbed
state is also usually strongly favored, $\epsilon \gg kT$.}
\label{well_barrier}      
\end{figure}

Physisorption and chemisorption will be carefully distinguished.
These are characterized by very different values of the local
monomer-surface association rate, $Q$ (see \fref{well_barrier}).  In
physisorption, monomer attachment is usually essentially
diffusion-limited, $Q=C/\ta$, where $\ta$ is monomer relaxation time
and $C$ is a system-dependent constant of order unity \cite{ben:fred}.
Chemisorption is normally much slower
\cite{ben:fred_letter,ben:fred,ben:grosberg_book} with $Q$ values
typically 8 or more orders of magnitude smaller than those of
physisorption.  The origin of this difference is that chemical bond
formation usually involves a large activation barrier (see
\fref{well_barrier}).  Similarly, desorption rates after chemisorption
are usually very small and can be ignored.  The effect of desorption
on physisorbing systems is more subtle and is discussed in section 4.
The above two classes naturally lead to very different adsorption
kinetics.  This is analogous to {\em bulk} polymer-polymer reaction
kinetics where depending on $Q$, polymer length $N$, and solvent
conditions, the kinetics are described by one of a range of
``diffusion-controlled'' and ``mean-field'' kinetic regimes
\cite{ben:grosberg_book,ben:reactiface_fund_letter,ben:reactiface_fund,ben:bulk_letter}.
Such regimes also arise for end-adsorbing polymers
\cite{ligoureleibler:end_funct_pol_interface,kramer:end_grafting_kinetics,
ben:reactiface_pol_letter,fredrickson:reactiface_prl,ben:reactiface_pol,
fredricksonmilner:reactiface_timedept,ben:fred_letter,ben:fred}.

In section 2 we briefly review the equilibrium picture for dilute
solutions and in section 3 we discuss experimental evidence for
non-equilibrium departures from this picture.  Theoretical work
related to physisorbing non-equilibrium layers from dilute solution is
reviewed in section 4.  We move to chemisorption, again from dilute
solution, in section 5.  Section 6 addresses irreversibility effects
involving melts and semi-dilute solutions.  We conclude with a brief
discussion of the experimental and theoretical outlook.

\setcounter{footnote}{0}
\section{Adsorption from Dilute Solutions: The Equilibrium Picture}

\subsection{Structure of Equilibrium Layers}

This section briefly outlines some central results of equilibrium
theories of adsorbed polymer layers.  Much more extensive reviews can
be found in refs.
\cite{fleer:pol_iface_book,netzandelman:polads_eqm_review_physrep_2003,
netzandelman:polads_eqm_review_2001,
clementjohner:polads_review_2000,eisenriegler:pol_surface_book}.  In
the scaling picture developed mainly in the 1980's
\cite{gennes:self_similar_profile,gennes:pols_interface_review,%
eisenriegler:proximal_exponent,eisenriegler:pol_adsorption_scaling_monte_carlo,%
gennespincus:proximal_exponent}, each adsorbed chain consists of
surface-bound monomers and large loops and tails generating a monomer
density profile $c(z)$ as a function of distance from the surface,
$z$.  Eisenriegler et al. \cite{gennespincus:proximal_exponent,
eisenriegler:proximal_exponent,eisenriegler:pol_adsorption_scaling_monte_carlo}
showed that very close to the surface, in the ``proximal'' region, the
density decays as a power law, $c(z) \twid z^{-m}$, where the critical
exponent $m \approx 1/3$ represents competion between surface-sticking
energy gain, chain entropy, and excluded volume interactions.  The
proximal region decay law crosses over to de Gennes' ``self-similar
grid'' regime
\cite{gennes:self_similar_profile,gennespincus:proximal_exponent,%
gennes:pols_interface_review}, $c(z) \twid z^{-4/3}$, beyond a certain
distance $\hprox$.  For $z>\hprox$ the polymer layer can be thought of
as a semi-dilute solution with continously varying local concentration
$c(z)$.  In this region the correlation length measuring the range of
excluded-volume interactions, $\xi = a^{-5/4} c^{-3/4}$, is
proportional to the distance from the surface, $z$, since this is the
only relevant length scale: $\xi \approx z$.  Here $a$ is defined to
be the monomer size.  Expressing $\xi$ in terms of $c$ leads to
\footnote{The cross-over distance $\hprox$ and the prefactor in the
proximal region density law can be determined by demanding (i) a
smooth cross-over at $\hprox$ and (ii) the osmotic free energy per
unit area, $\int_a^{\RF} dz kT/\xi^3$, balances the sticking free
energy per unit area, $\epsilon\, a\, c(a)$.}
                                                \begin{eq}{equilibrium}
\fl
a^3 c(z) 
\approx 
\casesshortii
{(a/\hprox)\, (a/z)^{1/3}} {a < z < \hprox }
{(a/z)^{4/3}}          {\hprox < z < \RF}
\gap
\hprox= a\, kT/\epsilon,\ \RF = a N^{3/5}
                                                                \end{eq}
Unless the bulk polymer concentration, $c$, is extremely small
\cite{bouchauddaoud:ads_pol_concentration_effects}, then the
equilibrium layer height is of order the Flory bulk coil radius $\RF$
as indicated in \eref{equilibrium}.  In this same range of $c$ the
adsorption isotherm exhibits a large plateau, \ie surface coverage
$\Gamma$ is weakly dependent on $c$.

Even in weakly adsorbing polymer systems, \eg adosrption through weak
van der Waals interactions, the value of $\epsilon$ is usually of
order $k T$.  By studying the adsorption/desorption transition in
binary solvent mictures, van der Beek et al.
\cite{vanderbeek:segmental_polads_energies_91} estimated the sticking
energies per monomer of many typical flexible polymers onto silica and
alumina surfaces from organic solvents to lie in the range $0.5$ to
6$kT$.  Hence the width of the proximal region is typically of order
the monomer size, $\hprox \approx a$, and a clear signature of the
proximal region is hard to probe experimentally.  In the following
we consider $\epsilon$ of order $kT$ or larger.  We remark that the
net monomer free energy of adsorption $\epsilon$ includes both the
``stickiness'' a monomer feels for the surface, but also the entropic
disadvantage due to constraining local orientational degrees of
freedom upon contact with the surface.  Thus, crudely speaking one can
say the stickiness contribution must exceed a crtical value
$\epsilonc$ representing the entropic disadvantage before a monomer
can adorb.  Computer simulations show $\epsilonc$ is of order $kT$ and
is lattice-dependent
\cite{eisenriegler:pol_adsorption_scaling_monte_carlo}.  The real
situation is more complex, with various contributions from electronic
and structural factors such as solvent molecule entropy effects,
etc\cite{israelachvili:book}.

The density decay law of \eref{equilibrium} reflects a power law
distribution of loop and tail sizes.  Neglecting differences between
loops and tails and the details associated with the proximal region,
then the loop size distribution per surface site is
\cite{gennes:ads_pol_loops,aubouy:scaling_flat_layers,semenovjoanny:loops_tails_europhys}
                                                \begin{eq}{input}
\Omega(s) \approx a^{-2} s^{-11/5} \period
\end{eq}
Beyond this, Semenov and Joanny
\cite{semenovjoanny:loops_tails_europhys} showed that the inner region
of the layer, $z < z^* \equiv aN^{1/2}$, is dominated by loops while
the outer region, $z > z^*$, is dominated by tails; the resulting
density profile obeys a $z^{-4/3}$ law above and below $z^*$,
respectively, but with different numerical prefactors.  Support for
the scaling conclusions of equations \eqref{equilibrium} and \eqref{input}
is provided by Monte-Carlo simulations of Zajac and Chakrabarti
\cite{zajacchakrabarti:pol_ads_statics_dynamics_monte_carlo}, de
Joannis et al.
\cite{dejoannis:polads_monte_carlo,dejoannis:polads_monte_carlo_europhysics},
and Cifra \cite{cifra:polads_monte_carlo_03}. These produce a density
decay consistent with the $z^{-4/3}$ law for long chains.  Zajac and
Chakrabarti
\cite{zajacchakrabarti:pol_ads_statics_dynamics_monte_carlo}
additionally report agreement with \eref{input}.

Complementary to the scaling approach outlined above has been the
numerical lattice model of Scheutjens and Fleer (SF)
\cite{scheutjensfleer:mean_field_adsorbed_pol_1,
scheutjensfleer:mean_field_adsorbed_pol_2}.  This is a self-consistent
mean field theory which averages excluded volume interactions and thus
treats self-avoidance in an approximate manner.  This approximation
however allows numerical solutions for the density profile and for the
loop and tail distributions and can additionally describe chains of
finite length.  The mean field description becomes more accurate for
solvents near the theta temperature (a common case)
where self-avoidance is
a weak perturbation except for the longest loops and tails.  The
existence of the loop- and tail-dominated regions of the layer was in
fact first established by the SF model
\cite{scheutjens:polads_tale_tails_86,scheutjensfleer:mean_field_adsorbed_pol_2}.
The layer height in the SF theory scales as $h \twid N^{1/2}$
\cite{scheutjensfleer:mean_field_adsorbed_pol_1,
scheutjensfleer:mean_field_adsorbed_pol_2} while the density profile
decays as $c(z) \twid z^{-2}$ (for sufficiently long chains),
different to the $z^{-4/3}$ decay predicted by the scaling approach,
as shown by van der Linden and Leermakers
\cite{vanderlinden:polads_mf_powerlaw_macro_92}.  {\em Analytical}
mean field solutions for the density profile in the limit of very long
chains were derived within the ground-state dominance approximation
\cite{gennes:book} by Jones and Richmond
\cite{jonesrichmont:mean_field_adsorbed_pol}.  Going beyond the ground
state dominance approximation, Semenov et
al. \cite{semenov:loops_tails_macro} subsequently generalized this
approach to account for finite length effects to leading order and
analytically describe the different contributions of loops and tails
to the density profile.  They found that loops dominate for $z <
z^*_{\rm MF} \equiv a N^{1/3}$ while tails dominate for $z > z^*_{\rm
MF}$, similarly to the scaling approach of Semenov and Joanny
\cite{semenovjoanny:loops_tails_europhys}.  These new methods have
revived interest in analytical and numerical mean field approaches to
polymer adsorption
\cite{johner:pol_ads_interpretation_numerical_results,
fleer:polads_analytical_mf_1_macro_99,fleer:polads_analytical_mf_2_macro_99,
blokhuis:week_polads_mf_formalism,avalos:polads_meanfield_1and2_macro_04,
ploehn:polads_mf_analytical_macro_1_88,ploehn:polads_mf_analytical_macro_2_89}.

Turning now to experiment, the fundamentals of polymer adsorption at
the solid/liquid and air/liquid interface have been studied in a vast
number of experiments.  Research prior to 1993 is reviewed in the book
by Fleer et al.  \cite{fleer:pol_iface_book}.  Given the strong
evidence for nonequilibrium effects (see below), in general one should
be very cautious when comparing experimental findings to equilibrium
expectations.  Overall, experiment is consistent with the general
trend predicted by equilibrium theories regarding structure of the
polymer layers which were studied, although the fine differences
between the mean field and scaling pictures are hard to distinguish.
Very briefly, measurements of the layer's surface bound monomer
fraction as a function of total adsorbed amount and molecular weight
(MW) by techniques such as NMR
\cite{barnett:nmr_pol_ads_macro_1981,beek:nmr_pol_ads_langmuir_1991},
ESR \cite{kobayashi:polads_esr_pmma_ps_90,robbsmith:esr_pol_ads}, or
infrared spectroscopy \cite{kawaguchi:polads_ps_ir_polymerj_80} give
results fairly consistent with the predictions of the SF theory
\cite{fleer:pol_iface_book}.  The thickness, $\hexp$, of polymer
layers has been probed as a function of chain length by hydrodynamic
methods \cite{fleer:pol_iface_book,
kawaguchi:exp_hydro_thickness_flow,%
cosgrove:exp_hydro_thickness,kato:exp_hydro_thickness_pcs},
ellipsometry \cite{kawaguchitakahashi:ads_pol_ellipsometry}, and the
surface force apparatus
\cite{kleinluckham:pol_ads_thickness_force_exp}.  Depending on the
method, $\hexp$ is proportional to a certain moment of the density
profile and many existent measurements are compatible with a power
law, $\hexp \twid N^\alpha$.  Certain studies have favored the SF
theory predictions
\cite{fleer:pol_iface_book,cosgrove:exp_hydro_thickness} while others
support the scaling predictions
\cite{kato:exp_hydro_thickness_pcs,kleinluckham:pol_ads_thickness_force_exp}.
For flexible polymer species the total surface coverage $\Gamma$ as a
function of bulk concentration is found to be very weakly dependent on
the polymer concentration in the bulk except for very dilute
solutions, in qualitative agreement with both the scaling and the SF
theories \cite{fleer:pol_iface_book}.  For a given bulk concentration,
meaurements of $\Gamma$ as a function of $N$ in good solvents
typically show a weak dependence on chain length for large $N$
\cite{fleer:pol_iface_book}.  This is consistent with the SF and
scaling theories which predict $\Gamma \twid \int_a^h dz \, c(z)$ is
dominated by the lower, $N$-independent limit.  

Small angle neutron scattering (SANS) and neutron reflectivity have
been used to probe the density profile.  These experiments established
the existence of a diffuse extended layer but at present there is no
general agreement as to the exact form of the density decay.  A
techical difficulty intrinsic to SANS, as dicussed in ref.
\cite{fleer:pol_iface_book}, is its limited sensitivity to the more
dilute regions of the layer.  Neutron reflectivity experiments are
also often ambiguous, since multiple density profiles can be fitted to
the same data.  The SANS experiments of Auvray and Cotton
\cite{auvraycotton:neutron_scattering_adsorbed_pol} and Hone et al.
\cite{hone:cosgrove_neut_scat_peo} are consistent with the $z^{-4/3}$
scaling law.  However the results of Hone et al. could also be
described by an exponential profile (see also
\cite{cosgrove:neutron_scattering_ads_pol_1987_macro}).
SANS experiments
by Cosgrove et al. \cite{cosgrove:neut_refl_block_homo_mica,
cosgrove:pol_ads_profiles_review_1990} do not support the scaling
predictions, but are more consistent with the SF theory.  Other SANS
studies by Rennie et al.  \cite{rennie:neut_refl_peo_air_water} are
inconsistent with both scaling and SF predictions, while neutron
reflectivity experiments of Lee et al.
\cite{lee:neutron_reflectivity_adsorbed_pol,guiselin:neut_refl_ads_pol_jcp}
and Sun et al. \cite{sun:pol_profile_neutron_reflection} have
generated data consistent with the scaling predictions.

\subsection{Single Chain Statistics and the
Equilibrium Distribution of Bound Fractions}

So far this section has discussed many-chain layer properties.
Equally important, and characteristic of the layer-forming processes,
are properties of {\em individual} surface-adsorbed chains in the
layer.  What is the spectrum of configurations of individual chains?
According to the scaling picture, a typical chain has $N D(s)$ loops
of length $s$ or greater, where $D(s) \equiv \int_s^\infty ds'
\Omega(s') \twid s^{-6/5}$ after using \eref{input}.  Semenov and
Joanny \cite{semenovjoanny:kinetics_adsorption_rouse} argue that
because of screening effects these are essentially independent blobs
and their 2D spatial extent parallel to the surface is $ [N
D(s)]^{1/2} a s^{3/5} = a N^{1/2}$.  This occurs for all scales $s$;
in particular, a given chain has of order one loop of length
$N^{5/6}$, also of size $aN^{1/2}$.  Hence a typical chain has a
lateral size of order $a N^{1/2}$, the ideal result (to within
logarithmic corrections
\cite{semenovjoanny:kinetics_adsorption_rouse}).

                                                   \begin{figure}[tb]
\centering
\includegraphics[width=6.8cm]{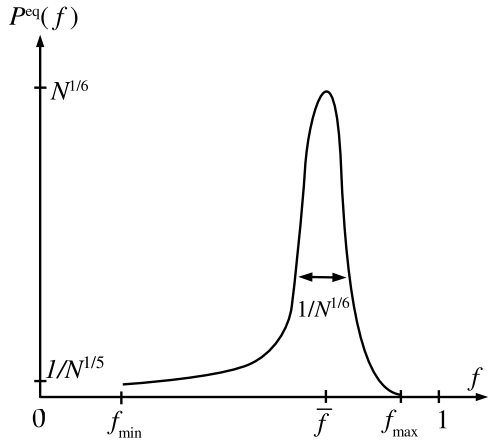}
\caption{\label{peq} Equilibrium probabilty distribution $\Peq$ 
of chain bound fraction, $f$,
in good solvents.  For very long chains the
distribution is sharply peaked at a value $\fbar$ of order unity.
For realistic values of $N$ the distribution is rather broad.  }
\end{figure}

A special role is played by another single chain property, directly
reflecting the degree to which individual chains are bound to the
surface.  This is the probability distribution $\Peq(f)$ that a chain
has a fraction $f$ of its monomers touching the surface.  This
property plays a central role in this review, since its features
closely reflect whether irreversible effects matter or not.  In two
independent Monte Carlo studies by Wang and Rajagopalan
\cite{wangrajagopalan:pol_ads_dynamics_monte_carlo} and Zajac and
Chakrabarti
\cite{zajacchakrabarti:pol_ads_statics_dynamics_monte_carlo} an
equilibrium distribution was found with a single peak at a value of
$f$ of order unity.  To our knowledge, $\Peq(f)$ has not been
calculated analytically, at least at the scaling level.  In order to
compare equilibrium predictions with experimental measurements of
bound fractions in non-equilibrium layers in later sections, we have
calculated $\Peq(f)$ by relating an adsorbed chain to a 1D
unidirectional walk combining an initial tail, a sequence of
intermediate steps, and a final tail.  The result, which is derived in
the appendix and is shown in \fref{peq}, reads:
                                                \begin{eq}{peq-peq}
\Peq(f) \approx {N^{-1/5} \over (\fmax - f)^{11/5}} 
\ \rho \square{(\fmax-f) N^{1/6}} \comma \gap
\rho[x] \gt \casesbracketsshortii{1}{x \gg 1}
                                 {0}{x \ll 1}
\end{eq}
This agrees qualitatively with the numerically measured distribution
of refs.  \cite{wangrajagopalan:pol_ads_dynamics_monte_carlo,
zajacchakrabarti:pol_ads_statics_dynamics_monte_carlo}.  Here
$\rho[x]$ is a scaling function which cuts off the distribution at a
small-scale-dependent parameter of order unity, $\fmax$.  The value of
$\fmax$ is determined by the small loops with length $\twid a$, whose
number is $\twid N$ per chain.  The mean value of the distribution,
$\fbar$ is located at $\twid \fmax - N^{-1/6}$.  The fluctuations
around the mean are due to the mass in long loops of length $N^{5/6}$.
There is order one such loop per chain and this diminishes $\fbar$ by
an amount of order $N^{-1/6}$.  \Eref{peq-peq} has a tail at
small $f$ which describes a small population of chains, a fraction of
order $N^{-1/5}$ of the total, with $f$ values far removed from the
mean: these are the chains with tails
\cite{semenovjoanny:loops_tails_europhys} of length up to order $N$
which determine the layer height $\approx a N^{3/5}$.  Notice in 
\fref{peq} we indicate that the distribution changes behavior at a
minimum value, $\fmin$, which is also of order unity (to within
logarithmic corrections in $N$
\cite{semenovjoanny:kinetics_adsorption_rouse}).  The value of $\fmin$
is determined by the largest loop or tail size in the layer,
expressions for which have been derived by Semenov and Joanny
\cite{semenovjoanny:kinetics_adsorption_rouse}.  We note that chains
with $f < \fmin$ values do exist but their number is predicted to be
much smaller \cite{semenovjoanny:kinetics_adsorption_rouse}.  For a
typical value $N = 1000$, one has $N^{-1/6} \approx 0.32$ and thus the
distribution is in practice rather broad.  Broadening effects in theta
solvents are expected to be even stronger.

\setcounter{footnote}{0}
\section{Dynamics of Adsorbed Polymers and Evidence 
for Non-Equilibrium Layers}

The most direct evidence for deviations from the equilibrium picture
derives from experiments on the kinetics of adsorbed polymers.  In
this section we review analytical and numerical studies of equilibrium
layer kinetics and then we proceed to review experiments showing
deviations from equilibrium.  Particularly revealing experiments have
been those probing bulk-layer exchange kinetics.

\subsection{Theories of Dynamics in Equilibrium Layers}

\begin{figure}[tb]
\centering
\includegraphics[width=12.5cm]{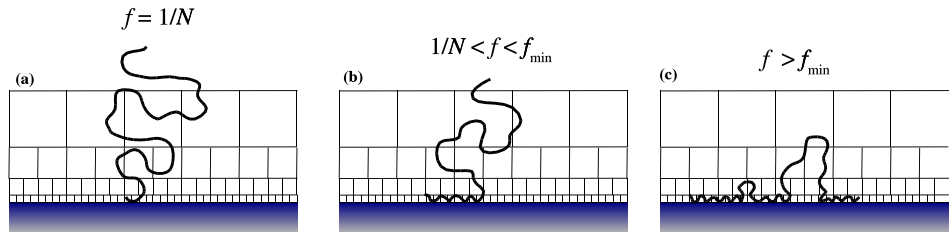}
\caption{\label{penetration} Schematic of new chain adsorption in an
equilibrium polymer layer (shown as a self-similar grid) as described
in ref. \cite{gennes:ads_pol_dynamics_new_trends}.  (a) Entry: a bulk
chain reptates into the layer and makes a first attachment to the
surface. (b) Spreading: the incoming chain establishes an increasing
number of contacts with the surface.  (c) A typical adsorbed chain
configuration is adopted, consisting of trains, loops and tails.  A
similar picture was used in
ref. \cite{semenovjoanny:kinetics_adsorption_rouse}.  Chain desorption
follows the same path in the reverse order.  }
\end{figure}

Compared to static properties, much less is established theoretically
about equilibrium dynamics.  These have been studied for good solvents
by de Gennes
\cite{gennes:ads_pol_dynamics_new_trends,gennes:ads_pol_dynamics_toyota,
gennes:penetration_layer,gennes:mobility_ads_chain,gennes:mobility_ads_chain_erratum},
Semenov and Joanny \cite{semenovjoanny:kinetics_adsorption_rouse}, and
Baschnagel et al. \cite{baschnagel:ads_kinetics_bidisperse} for
bidisperse solutions.  The picture emerging from these works is that
the layer of bound chains has a certain characteristic equilibration
time $\taueq$.  This can be thought of as the time after which the
chains following the distribution $\Peq(f)$ of \fref{peq} are
completely reshuffled among themselves.  The exchange of chains
between the bulk and the layer was predicted to be a slower process
due to the fact that incoming and outgoing chains have to pass through
unfavored configurations having a small fraction of bound monomers
(see \fref{penetration}).  de Gennes assumed reptation dynamics
(\ie entangled layers) and found the longest relaxation time of the
layer to scale as
\cite{gennes:ads_pol_dynamics_new_trends,gennes:mobility_ads_chain,%
gennes:mobility_ads_chain_erratum}
                                                \begin{eq}{reptation}
\taueq \approx \ts N^3  \gap {\rm (entangled\ layers).} 
\end{eq}
Here $\ts$ is the relaxation time of an adsorbed monomer which, due to
wall-polymer interactions, may be much larger than the corresponding
time in the bulk, $\ta$ \cite{zheng:poldiffusion_interface}.  Semenov
and Joanny \cite{semenovjoanny:kinetics_adsorption_rouse} assumed
unentangled layers and Rouse-Zimm dynamics and obtained $\taueq
\approx \ta N^2$ (in their work $\ts\approx\ta$ was assumed).

In equilibrium, layer and bulk chains desorb and adsorb at the same
average rate, respectively.  In de Gennes' picture bulk chains adsorb
in two stages (see \fref{penetration}).  During the first ``entry''
stage, the bulk chain overcomes the exclude-volume barrier presented
by the layer's loops and tails and makes its first contact with the
surface, \ie $f = 1/N$.  During a second ``spreading'' stage the chain
continues to make an increasing number of surface-contacts, $f$
increases up to $\fmin$, and the chain becomes part of the bound
layer.  When entry is rate-limiting he found that the mean lifetime of
an adsorbed chain before its desorption is $\tauex \approx \ta
N^{3.7}/ \phi$, where $\phi$ is the volume fraction of polymer in the
bulk.  Semenov and Joanny
\cite{semenovjoanny:kinetics_adsorption_rouse} described the dynamics
using a similar picture, but assuming unentangled layers and
Rouse-Zimm dynamics.  They obtained a slighlty different chain
lifetime, $\tauex \approx \ta N^{2.42}/\phi$ (to within logarithmic
corrections).

Note that the exchange timescale, $\tauex$, has a weak power law
dependence on $N$ rather than exponential because the
incoming/outgoing barrier is small.  The scaling $\tauex \twid 1/\phi$
reflects the linear dependence on concentration of the rate of chain
arrival at the surface.  Note also that even for the highest dilute
solution concentrations, $\phi = \phistar$, where $\phistar \equiv
N^{-4/5}$ is the chain overlap threshold concentration
\cite{gennes:book}, one still has $\taueq \ll \tauex$.  A prediction
of the above works is that chain desorption into pure solvent, $\phi
\gt 0$, is extremely slow, which is well-established experimentally
\cite{cohenstuartkeizer:polads_kinetics_review_2001}.

Now suppose one starts with a layer of labeled chains in equilibrium
and replaces the bulk solution with identical but unlabeled chains of
the same concentration at $t=0$.  An important prediction of the above
theories is that the decay of the surface coverage of labeled chains,
$\Gamma$, is a simple exponential for all times
\cite{gennes:ads_pol_dynamics_toyota,semenovjoanny:kinetics_adsorption_rouse}:
                                                \begin{eq}{exchange-simple}
\Gamma(t) = \Gamma(0) \ e^{-t/\tauex}
\end{eq}
An implicit feature of \eref{exchange-simple} is that there is a
single observed desorption rate since $\tauex \gg \taueq$, \ie the
desporption process is slow enough to sample an average over all
equilibrium chain states in the layer.  Note this result in fact
assumes a given desorbed labeled chain does not readsorb, \ie memory
is lost instantly.  Experimentally, this necessitates a mixing process
in the bulk to dilute the desorbed chains. In the absence of such
mixing diffusion returns a desorbed chain to the surface repeatedly,
an effect which can lead to non-exponential decay
\cite{ben:anomalous,wang:polads_exchange_kinetics_macro_95,ben:reactiface_fund}.

The kinetics of polymer layer build up starting from empty or
``starved'' surfaces is more complex and has been considered in refs.
\cite{gennes:ads_pol_dynamics_new_trends,gennes:ads_pol_dynamics_toyota,
hasegawadoi:scf_theory_ads_dynamics,semenovjoanny:kinetics_adsorption_rouse}.

\subsection{Dynamics of Adsorbed Polymer Layers: Monte Carlo Simulations}

This sub-section provides a brief review of numerical Monte Carlo
simulations of dynamics in many-chain polymer layers in contact with
dilute solutions (for dynamics of single chains see refs.
\cite{milchevbinder:single_ads_chain_dynamics_monte_carlo,
lai:polads_apparent_glassy_correct,liuchakrabarti:ads_single_moldyn,
ponomarev:artem_2d_diffusion,hernandez:polads_simul_single_hydrogen_energy_01}).
The simulations reported here did not include hydrodynamic
interactions.  The main results of the simulations by Wang et al.
\cite{wang:polads_exchange_kinetics_prl_95,wangrajagopalan:pol_ads_dynamics_monte_carlo}
are qualitatively in agreement with the theoretical picture of the
previous subsection.  They found that the lateral dynamics of adsorbed
chains up to $N=100$ are consistent with Rouse dynamics. For
sufficiently sticky surfaces ($0<\epsilon\le 1.5 kT$ with $\epsilonc =
0.5 kT$) the value of $\tauex$ was found to be much larger than the
lateral relaxation time, even though the scaling dependence on $N$ was
the same.  This should be contrasted with the Semenov and Joanny
prediction that the two exponents differ by a small value, 0.42.  Wang
et al. observed non-exponential exchange kinetics arising from
readsorption of desorbed chains.

Lai
\cite{lai:pol_ads_statics_dynamics_monte_carlo,lai:polads_apparent_glassy_correct}
studied the layer dynamics as a function of $\epsilon$ for $N \leq 80$
and interestingly found that for $\epsilon \gsim 1 kT$ (with
$\epsilonc \approx 0.9 kT$) the lateral chain dynamics started to slow
down and to approach apparently glassy dynamics at $\epsilon \approx 4
kT$.  This result was claimed to be valid despite an omission in the
implemented algorithm \cite{lai:polads_apparent_glassy_correct,
lai:single_ads_chain_statics_dynamics_monte_carlo}.  This report is
important since it indicates that the value of $\epsilon$ is crucial in
polymer adsorption.

Zajac and Chakrabarti
\cite{zajacchakrabarti:irrev_pol_ads_monte_carlo} studied the dynamics
for $N=100$ and $N=200$ and $\epsilon + \epsilonc = 1.8 kT$ near and
above $\phi = \phistar$.  Their algorithm involved unphysical
reptation moves in order to speed up the dynamics.  In equilibrium
they found a distribution of bound fractions similar to the one of
\fref{peq} and observed that the internal dynamics of reshuffling
of chains between different $f$ values is complex.  The timescale for
the exchange of adsorbed chains by bulk chains was found to be slower
than internal equilibration processes.  Simple exponential exchange
kinetics were observed as in \eref{exchange-simple}.

Takeuchi \cite{takeuchi:polads_exchange_simul_99} also used the Monte
Carlo method with extra reptation moves.  For surfaces with $\epsilon
\approx 1.6 kT$ he observed exponential exchange kinetics while for
$\epsilon \approx 0.9 kT $ readsorption effects were important leading
to non-exponential exchange kinetics.

\setcounter{footnote}{0}
\subsection{Experiment: Departure from Equilibrium Picture}

The first experimental studies of exchange kinetics on flexible
polymers were performed by Pefferkon et al.
\cite{pefferkon:ads_pol_dynamics_bimolecular,pefferkon:ads_pol_exhange_kinetics,%
varoqui:mobility_ads_pol_review,pefferkorn:polyacrylamide_ads_review_99}
using radioactive labeling techniques.  One study
\cite{pefferkon:ads_pol_dynamics_bimolecular,varoqui:mobility_ads_pol_review,
pefferkorn:polyacrylamide_ads_review_99} involved labeled
polyacrylamide (PAM) in water adsorbed through hydrogen bonding onto
aluminol-grafted glass beads. The beads were exposed to a dilute
solution of labeled PAM for approximately 30min until time-independent
coverage was achieved.
The labeled solution was then replaced by a dilute unlabeled PAM
solution of varying concentration $c$ and the amount of labeled PAM
which remained adsorbed, $\Gamma$, was monitored as a function of time
as shown in \fref{pefferkorn_pam}(i).  An interesting result of this
experiment was that the exchange rate per labeled chain, shown in
\fref{pefferkorn_pam}(ii), was {\em time-dependent} and reached a
constant value after a cross-over period of $\approx 300$min which was
approximately the same for every $c$.  This asymptotic rate was
found to increase linearly with $c$, as shown in
\fref{pefferkorn_pam}(iii).  The observed spectrum of exchange times
disagrees with \eref{exchange-simple} and this can be
interpreted in many ways as follows:
\begin{enumerate}

\item 

The observed non-exponential exchange kinetics is a signature of
non-equilibrium.  Pefferkorn et al.
\cite{pefferkon:ads_pol_dynamics_bimolecular,varoqui:mobility_ads_pol_review}
argued that the interface is populated with a spectrum of different
frozen or slowly changing configurations and, consequently, different
kinetic properties. (They proposed that the layer consists of a flat
sublayer of tightly bound chains which exchange slowly, plus a less
tightly bound population which exchange more rapidly).  

                                                   \begin{figure}[tb]
\centering
\includegraphics[width=8.9cm]{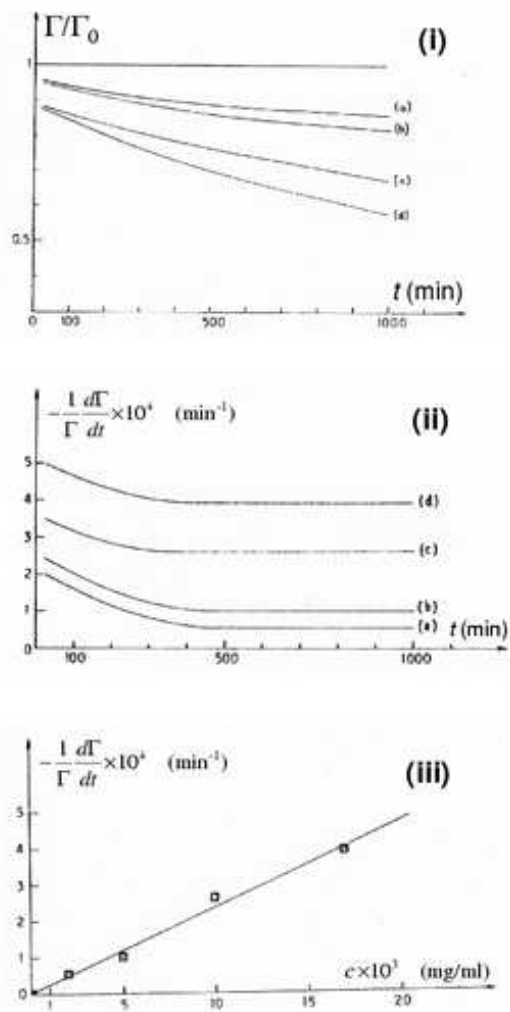}
\caption{\label{pefferkorn_pam} (i) Time-dependence of surface
coverage $\Gamma$ of radio-labeled PAM on glass beads in water in the
presence of polymer solutions at concentration $c = 2 \times
10^{-3}$ (a), $5 \times 10^{-3}$ (b), $10 \times 10^{-3}$ (c), $17
\times 10^{-3}$ mg/mL (d).  Unlabeled curve at top is desorption in
the presence of pure solvent. $\Gamma_0$ is initial surface
coverage. (ii) Specific exchange rate of radio-labeled PAM, from
surface to solution, as a function of time for different bulk
concentrations as in (i).  The rate is time-dependent which indicates
non-exponential exchange kinetics.  (iii) Asymptotic apecific exchange
rate of radio-labeled PAM as a function of bulk concentration.
[Reprinted with permission of John Wiley \& Sons, Inc. from Pefferkorn
E., Carroy A., and Varoqui R. 1985 {\em J.
Polym. Sci. Polym. Phys. Ed.} {\bf 23} 1997, Copyright \copyright 1985
John Wiley \& Sons, Inc]}
\end{figure}

\item 

The layer is not in equilibrium when the exchange experiment starts
but it equilibrates after $\taueq \approx$300min which is larger than
the layer's preparationg time.  The asymptotic exchange rate then
becomes constant and equal to $1/\tauex$.  The fact that
asymptotically $\tauex \twid 1/c$ as seen in
\fref{pefferkorn_pam}(iii) and the fact that $\tauex > 300$min as can
be seen in \fref{pefferkorn_pam}(ii), are consistent with this
interpretation and the theories reviewed in subsection 3.1.  Assuming
reptation dynamics, \eref{reptation}, and given $N \approx
1400$, this implies a relaxation time of adsorbed monomers of order
$t_s \approx 10^{-5}$s.  This is much larger than monomer relaxation
times in the bulk, $\ta \approx 10^{-10}$s.

\item The layer is in fact in equilibrium but its exchange kinetics
and internal equilibration processes are much more complex than
assumed by existent theories, at least for this system.  For example,
if the equilibrium $\Peq(f)$ is very broad and chains with different
$f$ values have very different exchange times, then the intial drop in
$\Gamma$ will be due mainly to the most rapidly desorbing chains if
their desorption times are less than $\taueq$. 

\end{enumerate}
Issues related to surface density of aluminol groups, polydispersity,
and effect of pH (this experiment was performed at pH=4 where PAM is
neutral while many of the surface aluminol groups were positively
charged
\cite{pefferkon:ads_pol_dynamics_bimolecular,pefferkon:ads_polbuild_up_1985})
may further complicate the dynamics.

\begin{figure}[htb]
\centering
\includegraphics[width=8cm]{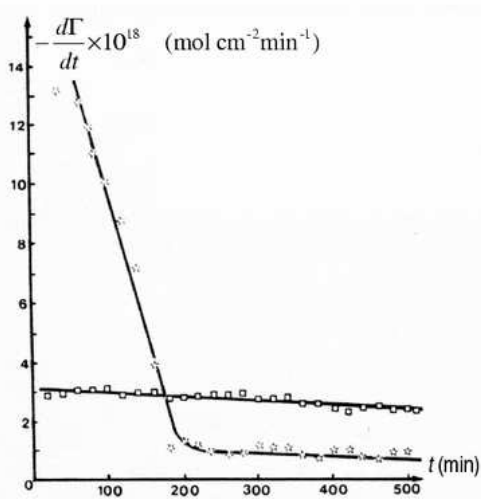}
\caption{\label{pefferkorn_ps}
Exchange rate of radio-labeled PS adsorbed on glass beads in CCl$_4$
as a function of time under different conditions.  Squares: $T=25^o$C,
$c = 0.9 \times 10^{-11}$mol/mL, initial surface coverage $\Gamma_0
= 3.4 \times 10^{-13}$mol/cm$^2$.  Stars: $T=35^o$C, $c = 5 \times
10^{-11}$mol/mL, $\Gamma_0 = 4.2 \times 10^{-13}$mol/cm$^2$.
[Reprinted with permission from Pefferkorn E., Haouam A. and Varoqui
R. 1989 {\em Macromolecules} {\bf 22} 2677.  Copyright 1989 American
Chemical Society.]}
\end{figure}

In refs.  \cite{pefferkon:ads_pol_exhange_kinetics,%
varoqui:mobility_ads_pol_review} Pefferkorn et al. performed an
experiment similar to their PAM studies, but for polystyrene (PS) in
CCl$_4$ (under good solvent conditions) adsorbing through hydrogen
bonds onto glass beads bearing surface silanol groups.  The measured
specific exchange rate of labeled chains shown in 
\fref{pefferkorn_ps} was found to be time-dependent for high surface
coverages, but independent of time for low $\Gamma$.  The
interpretation
\cite{pefferkon:ads_pol_exhange_kinetics,pefferkon:ads_pol_build_up_1988}
was that for small $\Gamma$ the layer consists of flattened-down
chains with identical exchange kinetics. At higher $\Gamma$ an outer
layer of loosely bound chains builds up and one returns to the
phenomenology of the PAM case (\fref{pefferkorn_pam}) with
time-dependent exchange rates.

\begin{figure}[t]
\centering
\vspace{.7cm}
\includegraphics[width=15cm]{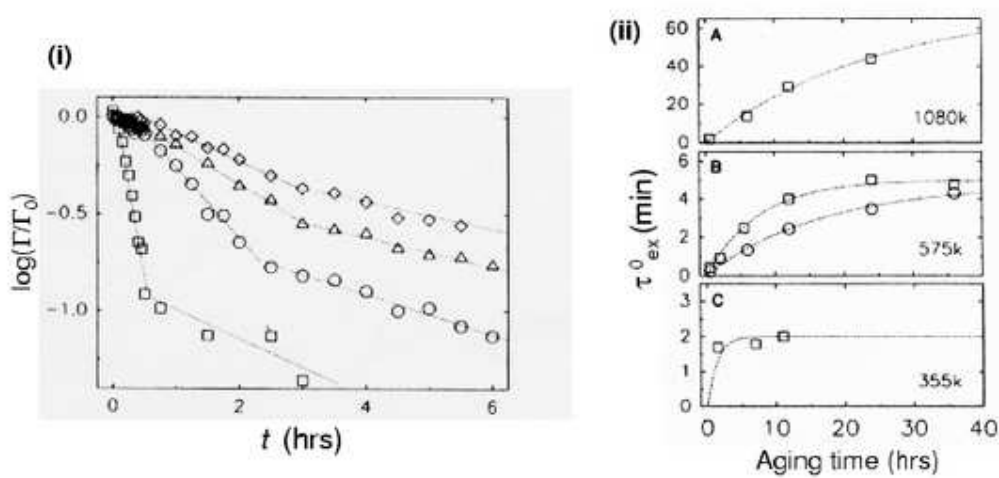}
\caption{\label{granick_ps_theta}
(i) Kinetics of exchange of the adsorbed mass of hPS adsorbed on
oxidized silicon in cyclohexane,
normalized by the surface
excess just before replacing hPS by dPS solution.
Different aging times are shown as follows.  Squares, 30min; circles,
6h; triangles, 12h; diamonds, 24h.  The initial decay of $\Gamma$ can
be fitted by an exponential, $\Gamma \twid e^{-t/\tauex^0}$.  (ii)
Initial exchange timescale, $\tauex^0$, plotted against aging times of
the initially adsorbed hPS of various molecular weights. A, MW =
1,080,000; B, MW = 575,000; C, MW = 355,000.  The MW of dPS is 550,000.
Squares and circles represent adsorption from solution concentrations
$c=1.0$ and 0.1 mg/mL, respectively.  [Reprinted with permission
from Frantz P. and Granick S.  1994 {\em Macromolecules} {\bf 27}
2553.  Copyright 1994 American Chemical Society.]  }
\end{figure}

Systematic experiments on PS exchange kinetics were also performed by
the group of Granick.  In a series of experiments, Frantz and Granick
\cite{frantzgranick:kinetics_adsorption_desorption_prl,%
frantzgranick:ps__cyclohexane_exchange_macro} studied the exchange of
adsorbed protonated PS (hPS) by bulk deuterated PS (dPS).
Measurements were performed in cyclohexane at the theta temperature of
dPS, 30$^o$C, slightly below the theta temperature of hPS, 34.5$^o$C
\cite{frantzgranick:ps__cyclohexane_exchange_macro}.  The attractive
surface was flat oxidized silicon.  Studying desorption in binary
solvent mictures (using a method developed by Cohen Stuart et al.
\cite{cohenstuart:polads_strength_displacer_84_1,
cohenstuart:polads_strength_displacer_84_2,
vanderbeek:segmental_polads_energies_91}) it was found that
\cite{frantzgranick:silicon_surface_preparation,vanderbeek:ps_on_silica_energy_89}
$\epsilon \approx 2 kT$.
After incubating the surface in dilute hPS solution for different
agings times, the solution was replaced by dilute dPS solution and the
adsorbed amounts of hPS and dPS was monitored as a function of time by
measuring infrared adsorption spectra (see \fref{granick_ps_theta}(i)).
Similarly to the experiments of Pefferkorn et al.
\cite{pefferkon:ads_pol_exhange_kinetics,%
varoqui:mobility_ads_pol_review} the exchange rate was time-dependent.
One sees in \fref{granick_ps_theta}(i) that at long times an
apparently exponential regime is reached as in
\fref{pefferkorn_pam}(ii).  The important result however is that the
time dependence at {\em short} times was found to be strongly
dependent on the layer's aging time.  The surface coverage of hPS at
short times could be fitted by an exponential law, $\Gamma \twid
e^{-t/\tauex^{0}}$.  It was found that $\tauex^{0}$ reached a constant
after aging times of order hours, depending on both $N$ (of order a few
thousand) and the concentration of the hPS solution (see 
\fref{granick_ps_theta}(ii)).  This suggests the layer structure was
contnuously evolving into an equilibrium or metastable state over
periods of hours.  The large aging time value of $\tauex^{0}$ was
found to be exponentially dependent on $N$ for a fixed molecular
weight of displacer dPS chains.

\setcounter{footnote}{0}

The results shown in figs. \ref{pefferkorn_pam} and
\ref{granick_ps_theta} are at least consistent with initially
non-equilibrium layers exhibiting a spectrum of exchange rates, which
then age towards equilibrium.  A possible model of these events can be
constructed as follows in terms of the chain-surface contact
distriution, $P(f)$.  According to theories
\cite{douglas:kinetics_pol_ads_jpc,ben:chemiphysi_letter,ben:chemiphysi_euro}
of the non-equilibrium layer, after adsorption onto a very sticky
surface $P(f)$ has a much broader distribution than the equilibrium
$\Peq(f)$.  This includes a finite fraction of chains with very small
contacts and fast bulk-surface exchange kinetics; it also includes a
fraction of order one which has the maxiumum possible number of
contacts, \ie chains lying fully adsorbed and flattened onto the
surface. One would expect these latter may have very large $\tauex$
values.  Thus when desporption is measured, the first $\tauex$ values
seen are those chains with small $f$ values; as time progresses larger
and larger $f$ values have had time to desorb. This is the $\tauex$
spectrum.  Only after the layer equilibration time $\taueq$ does the
distribution of $f$ values reach the equilbrium $\Peq(f)$ of
\eref{peq-peq}.  This is much less broad, centered on a mean $f$ value
corresponding to a $\tauex$ value much greater than $\taueq$.
Thereafter, a single averaged mean exchange time is observed.  It
should be mentioned, however, that in the experiments of
\fref{granick_ps_theta} this last phase of reaching equilibrium is not
manifest; it is possible that this system never reaches a true
equiibrium.

\begin{figure}[tb]
\centering
\includegraphics[width=15cm]{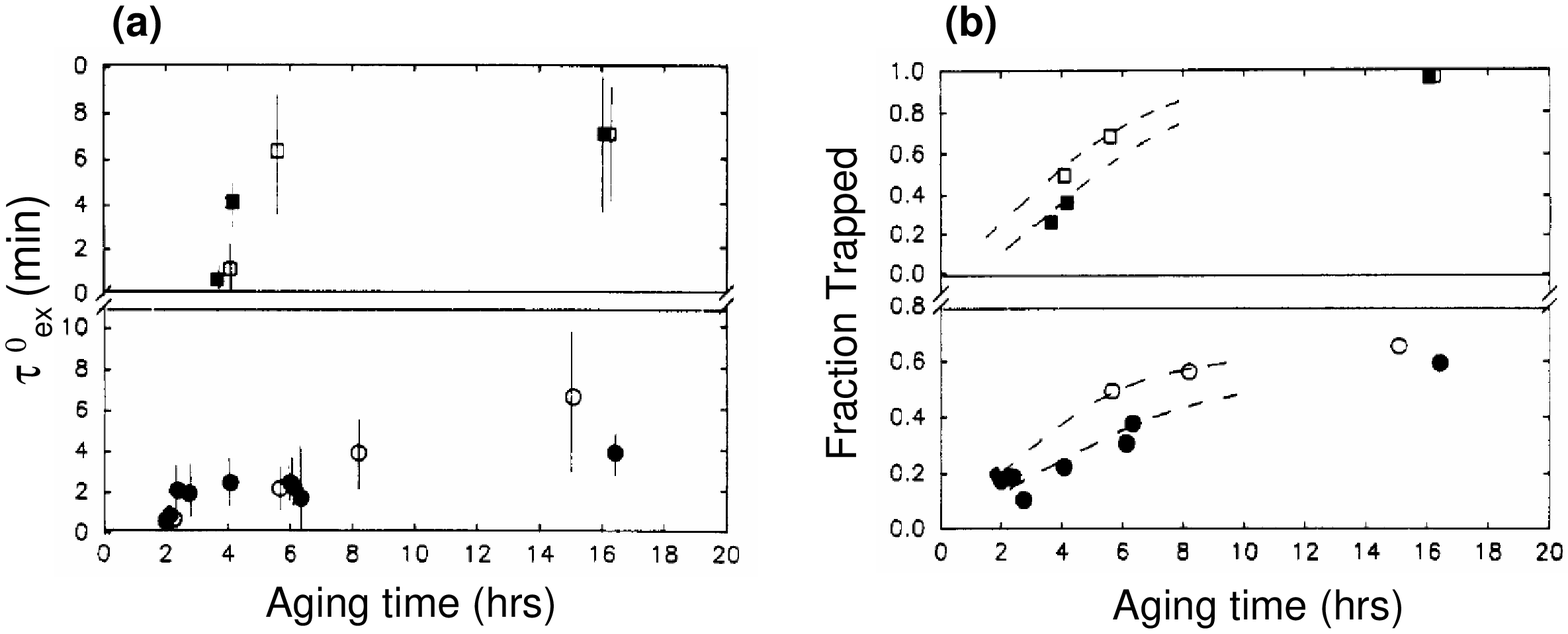}
\caption{\label{granick_ps_good}
(a) Plot of the initial exchange time constants, $\tauex^0$, for the
displacement of hPS adsorbed onto oxidized silicon in CCL$_4$ by dPS
versus aging time.  Bottom graph (circles): dPS MW=550,000.  Top graph
(squares): dPS MW=87,000.  Open (filled) symbols denote hPS MW=96400
(MW=575,000).  The diamond indicates a 96400 hPS/550,000 dPS exchange
following a 6h aging of a bare surface in pure solvent. 
(b) Trapped fractions of hPS plotted against aging times. Symbols as
in (a).
[Reprinted with permission
from Schneider H. M. and Granick S.  1992 {\em Macromolecules} {\bf 25}
5054.  Copyright 1992 American Chemical Society.]  
}
\end{figure}

In another study by Schneider and Granick
\cite{schneidergranick:kinetic_traps_exchange}, the displacement of
hPS by dPS adsorbed on oxidized silicon was explored in CCl$_4$ under
moderately good solvent conditions.  The adsorption energy was
$\epsilon \approx 1.3 kT$
\cite{schneidergranick:kinetic_traps_exchange,frantzgranick:silicon_surface_preparation,
vanderbeek:ps_on_silica_energy_89}, measured with the method of refs.
\cite{cohenstuart:polads_strength_displacer_84_1,
cohenstuart:polads_strength_displacer_84_2,
vanderbeek:segmental_polads_energies_91}.  Once again, $\tauex^0$ was
dependent on aging times, and it took hours for it to reach what might
be a plateau.  A main difference with respect to the theta solvent
case just described above is that at longer times of order hours, the
exchange rate became essentially zero and no significant exchange
could be detected over periods up to 10h (data not shown). Thus a
fraction of order unity of hPS chains remained trapped in the layers,
even when the displacer chains dPS chains were much longer than the
pre-adsorbed hPS (see \fref{granick_ps_good}(b)).  These findings seem
to support a picure where the evolution of the layer during aging
leads to a state containing more and more kinetically trapped chains.
This state was reached more rapidly for shorter chains and the trapped
fraction depended on the length of displacer chains.  The authors of
ref.  \cite{schneidergranick:kinetic_traps_exchange} argue that it is
unlikely that trapping occurs due to the slight difference between the
adsorption energies of hPS and dPS.  We note that Pefferkorn et al.
found initial exchange rates of order hrs$^{-1}$
\cite{pefferkon:ads_pol_exhange_kinetics} for a far more dilute system
involving PS in CCl$_4$ adsorbed onto silanol groups
(\fref{pefferkorn_ps}).

We note that experiments by Dijt et al.
\cite{dijt:pol_ads_exchange_peo_ps} on ``young'' PS layers formed in
flow from decalin solutions onto silica (at theta solvent conditions
where $\epsilon \approx 2kT$
\cite{dijt:pol_ads_exchange_peo_ps,vanderbeek:ps_on_silica_energy_89})
showed no evidence of significant non-equilibrium effects.  These
workers used optical reflectometry to measure the surface coverage
$\Gamma$ in mixtures of short (MW $\approx$ 10,000) and long (MW $>$
100,000) PS chains.  Unaged mixed polymer layers with different
compositions were generated which lead to measurable differences in
surface coverage.  It was found that layers with the same final
composition had approximately the same final surface coverage,
independently of adsorption history.  This was not true for transient
layers whose coverage however relaxed rather rapidly, in times of
order minutes.  At present it is not clear if and how the apparently
fast equilibration of $\Gamma$ is related to the relaxation of chain
configurations probed by exchange experiments using labeled chains
which appears much slower.

\begin{figure}[tb]
\centering
\includegraphics[width=13cm]{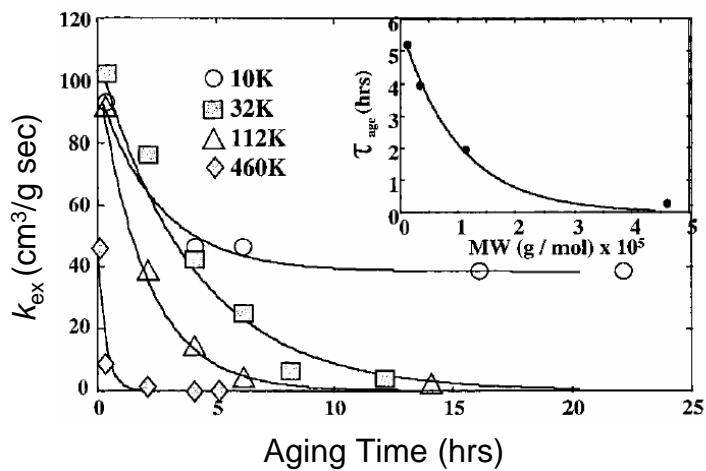}
\caption{\label{santore_peo}
Values of the initial exchange rate, $k_{\rm ex} \approx 1/(\tauex^0
c)$, for the displacement of labeled PEO chains adsorbed onto
silica in water by unlabeled chains of the same MW, as a function of
aging time and MW.  The curve through each data set is a single
exponential, $k_{\rm ex}(\tau) = k_{\rm ex} (0) e^{-t/\tau_{\rm
age}}$, summarizing the layer evolution for each particular molecular
weight.  The inset summarizes the aging time constants $\tau_{\rm
age}$ for the single-exponential fits in the main part of the figure.
[Reprinted with permission from Mubarekyan E. and Santore M. M.  2001
{\em Macromolecules} {\bf 34} 4978.  Copyright 2001 American Chemical
Society.]}
\end{figure}

Another widely studied adsorbing polymer is poly(ethylene
oxide) (PEO) for which the glass transition temperature (without
solvent) is $\Tg \approx -41^o$C and the melting temperature is $\Tm
\approx 70^o$C
\cite{brandrupimmergut:polymer_handbook99,mubarekyansantore:N_age_peo_exchange}.
PEO is thus considerably more flexible as compared to PS which has
$\Tg \approx 90^o$C \cite{brandrupimmergut:polymer_handbook99}.
Dijt et al.  \cite{dijt:pol_ads_exchange_peo_ps} studied mixed layers
of long and short PEO chains adsorbed onto silica via hydrogen bonding
in aqueous solutions using the same method they used for PS and found
fast relaxation dynamics for the coverage of unaged mixed layers.
More recently, Fu and Santore
\cite{fusantore:peo_competitive_adsorption,fusantore:age_relaxation_peo_exchange}
and Mubarekyan and Santore
\cite{mubarekyansantore:N_age_peo_exchange,%
mubarekyansantore:barrier_peo_exchange} have further explored the
kinetics of PEO adsorption using TIRF (total internal reflectance
fluorescence).  Their studies involved adsorption of unlabeled and
fluorescently labeled PEO onto silica from water, a good solvent, in
flow.  Microcalorimetry studies
\cite{trens:peg_silica_water_calorimetry} indicate $\epsilon \approx
1kT$.  It was found
\cite{fusantore:peo_competitive_adsorption,fusantore:age_relaxation_peo_exchange}
that the displacement of short PEO chains by long ones was fast for
unaged layers, qualitatively consistent with the findings of Dijt el
al.  \cite{dijt:pol_ads_exchange_peo_ps}.  However with increasing
chain length and aging, the exchange kinetics started to slow down and
could not be described by a single exponential exchange law as in
\eref{exchange-simple}
\cite{fusantore:age_relaxation_peo_exchange,mubarekyansantore:N_age_peo_exchange}.
For long chains a fraction of the layer appeared unexchangeable during
the experiment's duration, of order hours (this is similar to the
trapped fractions observed for PS by Schneider and Granick
\cite{schneidergranick:kinetic_traps_exchange},
\sref{granick_ps_good}(b)).  It was found that after a certain maximum
aging time further aging had no observable effect in the kinetics.
The observed aging effects are summarized in \fref{santore_peo}
where the initial exchange rate is plotted as a function of molecular
weight and aging time.  Aging in these experiments was performed in a
pure solvent rather than in contact with the dilute solution.  A
difference between the findings of the groups of Santore and Granick
(for PS) is that Mubarekyan and Santore report an initial exchange
rate reaching a plateau value more rapidly for longer rather than
short chains.  At long times the kinetics could be fitted by a
stretched exponential.  This may be due to a continuing relaxation of
the layer, but also they may be due to the effects of chain
polydispersity \cite{fusantore:age_relaxation_peo_exchange}.
Polydispersity effects may be very important in these systems due to
the tendency of long chains to displace the shorter ones, since long
chains suffer less entropic penalty when immobilized at the surface
\cite{baschnagel:ads_kinetics_bidisperse}.
In ref.  \cite{mubarekyansantore:barrier_peo_exchange} mixed layers of
labeled and unlabeled chains were formed and the kinetics of
fluorescent chain exchange with unfluorescent bulk chains was studied.
Provided the layers were aged for a period of a few hours, an
interesting result was that the kinetics were identical irrespective
of the adsorption history of the labeled subpopulation.  Thus the
picture of different classes of chains whose conformation is
determined by the time of their arrival on the surface appears
inapplicable for this system.

\begin{figure}[tb]
\centering
\includegraphics[width=14cm]{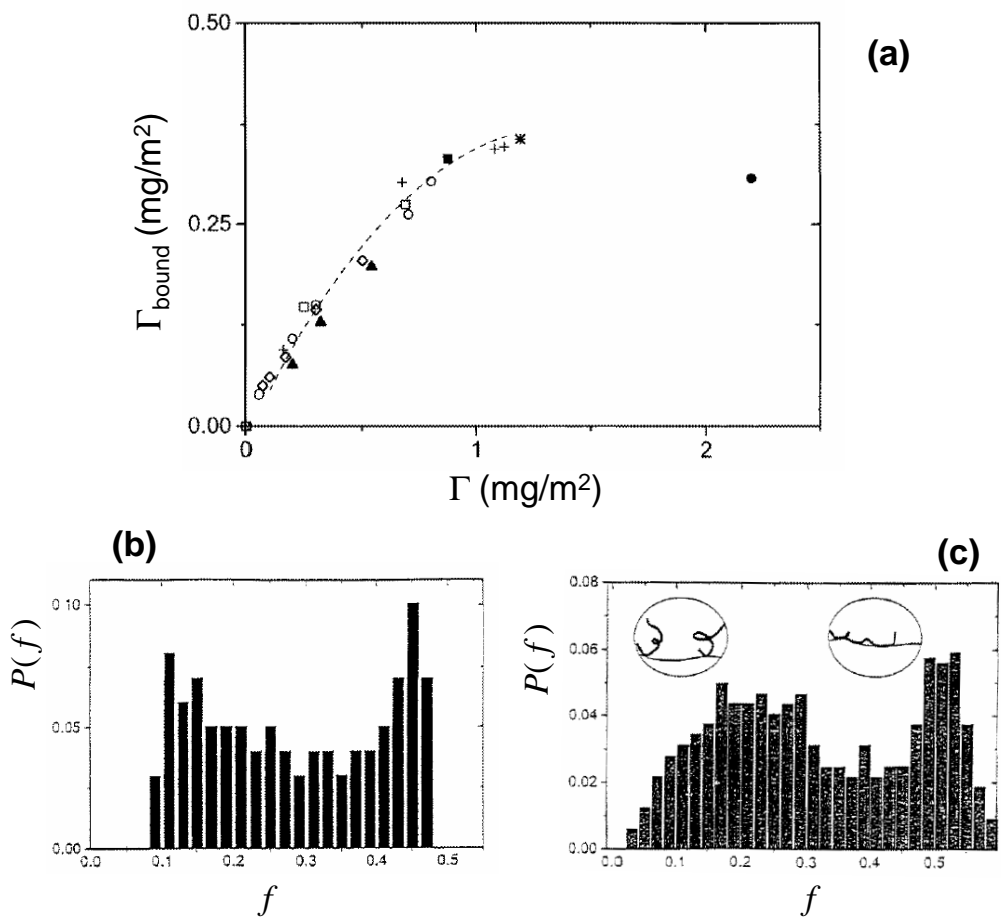}
\caption{\label{granick_pmma}
(a) Mass bound by hydrogen bonding to the surface, $\Gammabound$, versus
total mass adsorbed, $\Gamma$, for PMMA adsorption onto
oxidized silicon from CCl$_4$ at concentration $\phi = 0.01$mg/ml.
Dashed line: results from uninterupted adsorption experiments,
MW=107,000. Symbols: results from dosage experiments (see refs.
\cite{schneider:granick_bimodal,douglas:kinetics_pol_ads_jpc}).  Open
circles: MW = 90,000, diamonds: MW=107,000, filled squares: MW=64,000,
open squares: MW=7700, stars: MW=64,000 at $\phi=0.08$mg/ml, plus
signs: MW=107,000 in toluene solvent, filled circle: MW=107,000 from
semidilute solution. (b) Histogram of distribution $P(f)$ of 
chain bound fractions $f$, derived from uninterrupted adsorption
experiments shown in (a). (c) As in (b) but with $P(f)$ derived from
dosage experiments.
[Reprinted with permission from Douglas J. F., Schneider, H. M.,
Frantz P., Lipman R., and Granick S. 1997
{\em J. Phys.: Condens. Matter} {\bf 9} 7699.  Copyright 1997 IOP
Publishing Ltd]
}
\end{figure}

The most direct evidence of irreversible effects has been provided in
another series of experiments by Granick's group
\cite{schneider:granick_bimodal,douglas:kinetics_pol_ads_jpc,%
johnsongranick:exchange_kinetics_pmmma,frantzgranick:pmma_bound_fraction_early}
who followed the adsorption of protonated PMMA (hPMMA) and deuterated
PMMA (dPMMA) onto oxidized silicon via hydrogen bonding from a dilute
CCl$_4$ solution (pure PMMA has $\Tg \approx 100^o$C).  The
temperature was 30$^o$C, slightly above the theta temperature of
hPMMA, 27$^o$C \cite{johnsongranick:exchange_kinetics_pmmma}.  This is
a system with somewhat larger sticking energy, $\epsilon \approx 4 kT$
\cite{johnsongranick:exchange_kinetics_pmmma}.  These PMMA layers
exhibited extremely slow exchange rates with bulk chains: for layers
of hPMMA aged for $\approx 1$h and exposed to dPMMA solution,
$\tauex^0 \approx 100$h was measured, much longer than the duration of
the experiment \cite{johnsongranick:exchange_kinetics_pmmma}.
Infrared absorption spectroscopy was used to measure simultaneously
the total coverage $\Gamma$ and the surface-bound part $\Gammabound$
as they evolved in time
\cite{schneider:granick_bimodal,douglas:kinetics_pol_ads_jpc}.  This
revealed the functional relationship $\Gammabound(\Gamma)$, which was
found to be independent of molecular weight, see 
\fref{granick_pmma}(a).  Using deuterium labeling in combination with
infrared dichroism measurements they could measure the bound fraction
$f$ of chain subpopulations.  It was found that early arriving chains
had much higher $f$ values than late arrivers, and these $f$ values
were frozen in during the experiment's duration.  Measuring the
distribution of $f$ values among chains they found a bimodal
distribution shown in \fref{granick_pmma}(b),(c) with two peaks at
small and large $f$, respectively.  This is strikingly different to
equilibrium layers where the distribution exhibits a unique peak shown
in \fref{peq}; for very large $N$ all chains within an
equilibrium layer become statistically identical and to within small
fluctuations are characterized by the {\em same} value, $f =
\Gammabound/\Gamma$.  The results of other experiments by Soga and
Granick \cite{sogagranick:ir_dichroism_pdms} studying
poly(dimethylsiloxane) (PDMS) adsorbing onto oxidized germanium from
CCl$_4$ were consistent with the existence of a bimodal distribution.

Parallel to the above works on exchange kinetics, evidence for very
slow and possibly non-equilibrium behavior has been gathered in
experiments measuring forces between polymer covered surfaces using
the surface force apparatus.  These systems have been modeled using
equilibrium concepts
\cite{gennes:polads_force_layers_macro_82, rossipincus:starved_pol_layers_macro,kleinrossi:analysis_expts_polads_scaling,mendezalcaraz:irrev_layers_interaction_macro_98}.
Significant hysteresis effects are, however, common: for PS adsorption
onto mica in near-theta solvents, Hu and Granick
\cite{hugranick:surface_force_ps_theta_90} and Ruths et al.
\cite{ruths:compression_ps_layers_theta_97} have observed that
following the first compression run, the force as a function of
surface separation during decompression is different and corresponds
to layers of smaller thicknesses.  Ruths et al. observed that force
profiles upon recompression did not relax back to the original profile
for days.  These workers assigned the observed phenomenolgy to
extremely slow relaxation processes.  (For these systems the authors
estimate
\cite{ruths:compression_ps_layers_theta_97,granick:pams_mica_force_direct}
$\epsilon \gsim 0.3 kT$).  Similarly slow dynamics were observed by
Ruths et al. \cite{ruths:surface_force_pep_96} for
poly(ethylene-propylene) on lipid coated mica in good solvent.

For the more flexible PEO systems, hysteresis effects are also
observed but appear to relax faster.  Luckham and Klein
\cite{luckhamklein:surface_force_peo_ps} and Raviv et al.
\cite{raviv:surface_force_shear} studied PEO adsorbed onto mica in
toluene, a good solvent (in ref. \cite{raviv:polymer_mat} it was
estimated that $\epsilon \approx 0.25kT$ after using the theory of
ref.  \cite{kleinrossi:analysis_expts_polads_scaling} to relate
osmotic pressure to sticking energy).
The layers were incubated in solution overnight, the two surfaces
brought into contact, and the force measured as a function of
separation using a surface force balance.  The force profile upon
decompression or rapid subsequent recompression was found to be
different to the initial profile, describing a layer of a smaller
thickness than the one during the first compression.  Recovery of the
original force profile occured more rapidly for shorter polymers.
Recovery times for the longer polymers (molecular weight 310,000) was
of order an hour.  Theoretical work by Raviv et al.
\cite{raviv:polymer_mat} assuming irreversibly frozen chain
configurations of the compressed layer is consistent with the measured
force-distance relationship immediately after compression.  Johner and
Semenov \cite{johnersemenov:comment_polymer_mat} proposed that a dense
phase of PEO nucleates on the surface.  There is also evidence, albeit
controversial
\cite{duval:cluster_formation_peo_macro_00,porschsundelof:peo_apparent_aggregation_macro_95},
that PEO forms aggregates in solution which adsorb onto surfaces
\cite{boilshair:polads_peo_toluene_aggregates_93,
marrahair:surfaceforce_ps_effect_solvent_88}.  If so, this could
be an important complication in all PEO systems.

Further information on the dynamics of adsorbed polymers in solution
arises from experiments on systems which are by their nature far from
equilibrium.  This includes studies of the kinetic build up of layers
on initially bare surfaces.  A recent review of kinetics in
such systems is given by Cohen Stuart and de Keizer
\cite{cohenstuartkeizer:polads_kinetics_review_2001}.  Another class
of far from equilibrium systems involves the displacement kinetics of
adsorbed chains by chains of a different species.  Experiments
performed by the Granick group
\cite{johnson:topological_influences_adsorption_desorption_prl,%
johnsongranick:adsorption_dynamics,schneider:granick_ps_displace_isoprene_1,%
schneider:granick_ps_displace_isoprene_2} involved PS adsorbed onto
oxidized silica over a period of 1 hour in a good solvent (CCl$_4$),
displaced by a more strongly adsorbing polymer: Johnson et
al. \cite{johnson:topological_influences_adsorption_desorption_prl,%
johnsongranick:adsorption_dynamics} studied displacement by PMMA while
Schneider et al. \cite{schneider:granick_ps_displace_isoprene_1,%
schneider:granick_ps_displace_isoprene_2} studied displacement by
polyisoprene.  These studies found that coverage of surviving PS
chains, $\Gamma_{\rm PS}$, obeyed long time stretched exponential
laws, $\Gamma_{\rm PS} \twid e^{-(t/\tau')^{\beta}}$, where $\tau'$
was of order hours and strongly temperature dependent.
Dijt et al. \cite{dijt:pol_ads_exchange_different} studied
displacement of PS by poly(butyl methacrylate) (PBMA) and
poly(tetrahydrofuran) (PTHF), and displacement of PBMA by PTHF.
Adorption occured on oxidized silicon in decalin solutions.  Their
results suggested that chain flexibility plays important roles in
determining the magnitude of exchange timescales.  Finally, evidence
for slow relaxation processes in adsorbed polymer layers also comes
from ESR measurements by Pan et al.
\cite{pan:durning_turro_esr_macro_95} for adsorption of PS onto porous
silica in CCl$_4$.

\setcounter{footnote}{0}
\section{Theoretical Work on Non-Equilibrium Physisorption}

In the previous section we reviewed experimental evidence for
non-equilibrium effects in polymer layers in solution.  In summary it
appears that, with increasing aging, chains adsorbed on strongly
attracting surfaces undergo internal relaxation processes and become
progressively more attached to the surface and harder to remove.  This
internal relaxation process becomes slower for even more sticky
surfaces, and for systems such as PMMA on oxidised silicon the chains
appear to become effectively frozen into configurations which depend
on their adsorption history.  The universality of the existent
observations is unclear.  

In this section we review theoretical work addressing these issues.
Despite many proposals, the nature of the very slow dynamics is not
well understood.

\subsection{Single Chain Physisorption}

Perhaps the theoretically simplest problem to attack theoretically is
single-chain adsorption in the limit of complete irreversibility of
monomer-surface bonds.  For physisorption, each attachment is
typically very rapid after monomer-surface contact, requiring a small
number of monomer-surface collisions.  We are not aware of first
principles analytical theories addressing this problem which account
for the complex polymer hydrodynamics near an interface, subject to an
increasing number of surface-polymer bonding constraints.  (However,
somewhat related issues have been addressed theoretically in the
context of chain collapse kinetics in a poor solvent
\cite{gennes:collapse_kinetics}).  Ignoring hydrodynamics, one might
expect an upper bound on the adsorption timescale $\tauads$ to be the
bulk coil relaxation time $\taubulk$
\cite{gennes:book,doiedwards:book}, since the constraints formed by
attachments would presumably accelerate the rate of arrival of new
monomers onto the surface.  However, the validity of even this
statement is no longer clear when polymer-surface hydrodynamics are
important.

Numerical studies have provided important insight. Shaffer
\cite{shaffer:strong_ads_heteropolymers} used the bond fluctuation
model to study single chain adsorption on a surface with sticking
energy $10kT$.  Due to the high cost in computer time hydrodynamic
interactions were not included.  Since there was no activation barrier
for monomer adsorption, his results correspond physically to almost
irreversible physisorption.  Using $N \le $ 80, he found that in the
presence of excluded volume interactions (physically corresponding to
dilute solution) $\tauads \twid N^{1.58}$.  In the absence of both
excluded volume interactions and hydrodynamics (which would correspond
physically to adsorption of a few stick chains from a non-adsorbing
polymer melt matrix), $\tauads \twid N^{1.5}$.  Note that both
timescales are weaker powers of $N$ than the corresponding bulk
relaxation times \cite{shaffer:strong_ads_heteropolymers,gennes:book}.
Similar scalings to those of Shaffer were independently obtained by
Ponomarev et al.  \cite{ponomarev:artem_durning} who used the
bond-fluctuation model for $N \le 100$.  The kinetics of single chains
adsorbing onto strongly attractive surfaces have also been studied by
molecular dynamics simulations by van Eijk et al.
\cite{vaneijk:ads_spread_theory_moldyn} and Liu and Chakrabarti
\cite{liuchakrabarti:ads_single_moldyn}

Konstadinidis et al. \cite{konstadinidis:irrev_ads_monte_carlo} used
Monte-Carlo simulations to study the structure of a single fully and
irreversibly collapsed chain on a surface.  In the collapsed state,
whose nature was dependent on the adsorption kinetics, steric and
topological constraints at the monomer level prevented every chain
unit from actually touching the surface.  The fraction of a flattened
chain's monomers which are physisorbed to the surface was measured to
be $\omega \approx 0.85$.  This should be compared with the value that
this simulation would have produced, $\omega\gt 1$, if the system were
allowed to reach equilibrium by introducing, say, a very small
desorption rate. This is because in the limit $\epsilon \gg kT$ the
nature of this particular lattice model allows the chain to completely
flatten down into a two-dimensional ``pancake''
\cite{gennes:book,gennes:ads_pol_dynamics_toyota}.

An important open issue is how to go beyond complete irreversibility
and study, for very sticky surfaces, the dynamics of a single chain
after flattening down onto the surface.  These dynamics are relevant
to layer formation on sticky surfaces during the early stages when the
layer consists of isolated single chains.  In a simple picture, one
might expect that for $\epsilon > kT$ monomer relaxation rates are
locally slowed down by a Boltzmann factor $\twid e^{(\epsilon+u)/kT}$
associated with local kink formation laterally or normal to the
surface (here $u$ is the height of the adsorption potential barrier,
see \fref{well_barrier}(b)).  However, the scaling with respect to $N$
would remain a power law.
\footnote{For
weakly adsorbed chains, this scaling has been measured
\cite{sukhishvili:granick_pancake_diffusion_nature,%
sukhishvili:granick_2d_diffusion_macro,chul:granick_2d_diffusion_opinion,%
maier:dna_2d_diffusion_prl,maier:dna_2d_diffusion_macro} and
numerically simulated
\cite{milchevbinder:single_ads_chain_dynamics_monte_carlo,ponomarev:artem_2d_diffusion}.
In ref. \cite{sukhishvili:granick_pancake_diffusion_nature,%
sukhishvili:granick_2d_diffusion_macro} using poly(ethylene glycol)
dilutely adsorbed onto self-assembled hydrophobic monolayers, condensed
octadecyltriethoxysilane, with $\epsilon \approx 0.5 - 1 kT$, simple
power laws were measured for the chain's self-diffusivity: $D \twid
N^{-3/2}$.} 
On the other hand, dynamics may be more fundamentally modified both by
local steric and global topological constraints.  Due to local
constraints the degree to which a monomer in this flattened
configuration can move laterally or desorb is likely to be strongly
dependent on the adsorption state of nearby monomers.  Thus its motion
might necessitate the cooperative motion of its neighbors which
introduces kinetic barriers.  Chakraborty and coworkers
\cite{chakrabort:glassy_polymer_solid,%
chakrabortyadriani:glassy_polymer_solid_2,%
shafferchakraborty:pmma_chemisorption_kinetics,%
adrianichakraborty:kinetic_ising_chemisorption} studied numerically a
one-dimensional Ising model of the effect of such kinetic barriers.
For $\epsilon>kT$ stretched exponentials were found in the dynamics of
correlation functions.  Relaxation times increased with increasing
$\epsilon$ and in certain cases diverged as $\epsilon$ increased
towards a value of order $kT$.  These results are similar to those
otained in kinetic Ising models of glasses
\cite{fredricksonandersen:kinetic_ising_prl}.  They were intepreted to
indicate the onset of glassiness in strongly adsorbing systems.  The
effect of global topological constraints on the dynamics has not been
analyzed to our knowledge.

\subsection{Many-Chain Adsorption}

Although at present there is little theoretical work addressing the
precise nature of the extremely slow dynamics in adsorbed layers,
past work has suggested a number of potentially important factors as
follows. 

\begin{enumerate}

\item Adsorbed monomer relaxation times may be considerably lengthened
due to Boltzman factors $e^{(\epsilon + u)/kT}$ associated with
desorption (see fig. \ref{well_barrier}).

\item Topological constraints may be important, as suggested in refs.
\cite{granick:comment_polymer_mat,johnersemenov:comment_polymer_mat,
raviv:reply_comments_polymer_mat,mubarekyansantore:N_age_peo_exchange}.
It is natural to expect that if the equilibrium layer is dominated by
loops of order the monomer size, then mutual pinning down of chains
may become very restrictive.  Equilibrium theory does indeed predict
that when $\epsilon$ reaches values $\twid kT$ or greater most of the
mass of a typical chain belongs to such small loops.  This hints at a
possible transition to more glass-like behavior at this point.

\item Equilibrium theory predicts that for $\epsilon \ge kT$, the
monomer density at the surface ($z \approx a$) becomes of order the
density in a polymer melt, see \eref{equilibrium}.  Kremer
\cite{kremer:ads_pol_glassy} and de Gennes
\cite{gennes:ads_pol_dynamics_toyota} have suggested that if the
solution temperature is below $\Tg$ for the melt (as it would be, for
example, for PS and PMMA at room temperature), glassy effects might
onset at the interface where polymer densities are similar to those of
melts.  Interactions with the surface may further enhance glassiness.
Consistent with this picture, increases of glass transition
temperatures have been observed for thin polymer melt films deposited
onto strongly adsorbing surfaces
\cite{keddie:films_tg_faraday_94,jones:thin_films_review_99}.

\item de Gennes
\cite{gennes:phase_sep_second_type,gennes:water_soluble_special_features_92}
suggested that a dense polymer phase may form near the sticky surface
for water soluble polymers such as PEO.  This would block interfacial
chain motion. Johner and Semenov
\cite{johnersemenov:comment_polymer_mat} proposed an analogous
mechanism for PEO in an organic solvent but this was disputed by Raviv
et al.  \cite{raviv:reply_comments_polymer_mat}.  Analogous phenomena
might occur for other polymers which tend to form crystalline phases.
Chain stiffness may also enhance local chain alignment and surface
ordering effects.

\item The dynamics in thin polymer melt films are often much slower
than in the bulk, even above $\Tg$
\cite{jones:thin_films_review_99,zheng:poldiffusion_surface_long_range}.
The theory of Semenov \cite{semenov:2d_rep} and of Semenov and Johner
\cite{semenovjohner:notes_dense_layers} for confined polymer films
predicts relaxation times which depend exponentially on molecular
weight, similar to the activated reptation model of bulk entangled
melts
\cite{deutsch:3point4_letter,deutsch:3point4,semenovrubistein:activated_rep_euro}.
Unlike the unconfined case, Semenov found exponential dependence for
even short entangled polymers in the confined situation.  Semenov and
Joanny \cite{semenovjoanny:kinetics_adsorption_rouse} proposed that a
similar phenomenology might apply to adsorbed polymers in solution.

\end{enumerate}

Most theoretical studies of non-equilibrium effects have considered
the limit of complete irreversibility upon attachment.  This is
directly relevant to more sticky systems such as PMMA on oxidised
silicon \cite{schneider:granick_bimodal,douglas:kinetics_pol_ads_jpc}.
Barford et al. \cite{barford:irrev_ads_1} studied irreversible
physisorption, motivated by experiments on polyelectrolyte adsorption.
Since the limit of high salt concentration was considered, this
analysis is also relevant to neutral polymers.  The build up of the
polymer layer profile $c(z,\Gamma)$ was followed as a function of
increasing surface coverage $\Gamma$.  An incremental increase in
adsorbed amount $d\Gamma$ contributes a change in density $d
c(z,\Gamma)$ whose form was determined by numerical solution of
self-consistent mean-field equations in which the excluded-volume
repulsion due to earlier arriving chains was assumed to be represented
by a {\em fixed} external field.  Physically, this models a situation
where new chains are irreversibly frozen in after exploring the phase
space which was available upon their arrival.  The incremental
profiles were integrated up to the saturation surface coverage,
$\Gammainf$, to obtain the final profile $c(z,\Gammainf) =
\int_0^{\Gammainf} d c(z,\Gamma)$.
                                                   \begin{figure}[htb]
\centering \includegraphics[width=12.5cm]{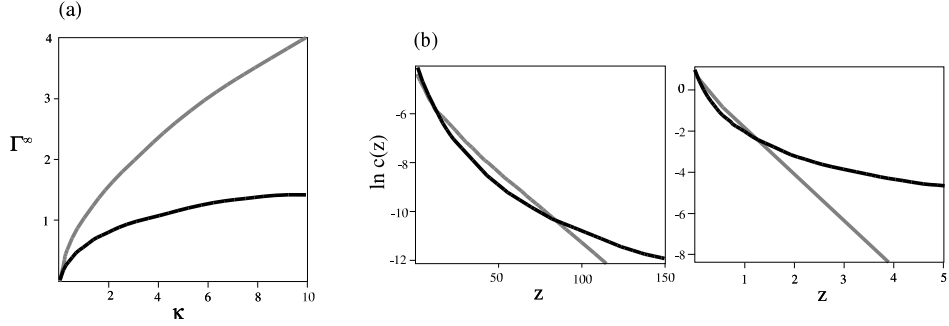}
\caption{\label{ball} Results of Barford et
al. \cite{barford:irrev_ads_1} for
irreversible adsorption (black lines) compared to equilibrium (gray
lines).  Scales are arbitrary.  (a) Total surface coverage $\Gammainf$
as a function of $\kappa$, a dimensionless measure of monomer sticking
energy. (b) Density profile $c$ as
a function of distance from surface.  
$\kappa = 0.1$ and $\kappa = 5$ shown on left and right, respectively.
Reproduced from ref. \cite{barford:irrev_ads_1}.
}
\end{figure}

The results of Barford et al. \cite{barford:irrev_ads_1} for
irreversible adsorption compared to equilibrium self-consistent
mean-field theory are illustrated in \fref{ball}.  The asymptotic
surface coverage was found to be less than the equilibrium value (see
\fref{ball}(a)).  Relative to the equilibrium density profile $c(z)$,
the irreversible profile was larger at small and large $z$ but smaller
for intermediate $z$ values (see \fref{ball}(b)).  Another prediction
is that the final profiles depend on sticking energies, though results
for very high sticking energies should be interpreted with caution
since the equilibrium self-consistent theory predicts $\Gammainf \gt
\infty$ for infinite chain-surface sticking energy (see
\fref{ball}(a)).  In a subsequent work by Barford and Ball
\cite{barfordball:irrev_ads_2} the condition of frozen density
profiles was relaxed and only the density at the interface was assumed
to remain frozen.  The final conclusions remained very similar.

Douglas et al. \cite{douglas:kinetics_pol_ads_jpc} modelled
irreversible adsorption by simulating a random sequential adsorption
process \cite{talbot:rsa_review}.  In this work chains were visualized
as deformable droplets.  Initially, when a chain arrives at a bare
surface, each droplet adsorbs onto a certain maximum cross-sectional
area.  As available surface area for adsorption become scarce, in
order for late-arriving chains to adsorb it was assumed droplets
deform by reducing their cross-sectional areas parallel to the surface
to fit into the empty space. In so doing, they become more extended
into the bulk.  Using this model a $P(f)$ distribution was generated
whose shape is similar to the experimental one of 
\fref{granick_pmma}(b).

\begin{figure}[tb]
\centering
\resizebox{0.8\columnwidth}{!}{%
\includegraphics{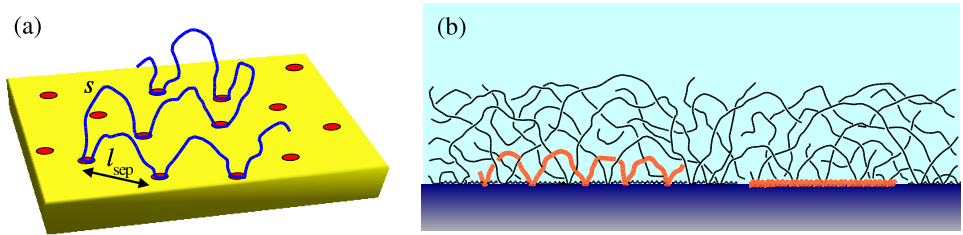} }
\caption{(a)
Typical configuration of a chain adsorbed during the late
stages, according to theory of refs.
\cite{ben:chemiphysi_letter,ben:chemiphysi_euro}.  Such chains can
adsorb onto free empty sites only (shown as discs) which are separated
by $\lsep$.  In so doing they form loops of $s$ monomers, with
$as^{3/5}=\lsep$.
(b) Sketch of final irreversibly formed layer structure,
as predicted in refs.
\cite{ben:chemiphysi_letter,ben:chemiphysi_euro}.  The layer consists
of two parts (one chain from each part is highlighted): (i) An inner
region of flattened chains making $\approx \omega N$ contacts per chain,
where $\omega$ is of order unity. (ii) A diffuse outer layer built up
from chains each making $f N \ll N$ contacts with the surface.  The
values of $f$ follow a distribution $P(f) \twid f^{-4/5}$.  Each $f$
value corresponds to a characteristic loop size for a given chain, $s
\approx \ncont/f$. 
}
\label{late_stages}      
\end{figure}

\begin{figure}[tb]
\centering
\includegraphics[width=10cm]{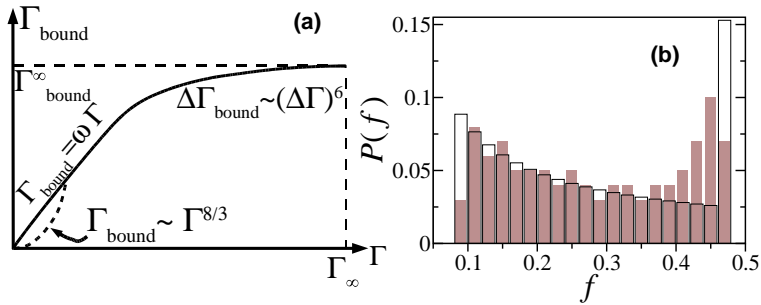}
\caption{\label{gamma_pf}
(a) Adsorbed polymer mass $\Gamma$ versus surface-bound part
$\Gammabound$ as predicted by refs.
\cite{ben:chemiphysi_letter,ben:chemiphysi_euro}.  For irreversible
physisorption, early arriving chains have a fraction $\omega$ of
their monomers bound to the surface and during early stages
$\Gammabound \approx \omega \Gamma$.  The early stage prediction for
chemisorption is different,
$\Gammabound\twid \Gamma^{8/3}$. The predicted form should be compared
with the experimental results shown in \fref{granick_pmma}(a).
(b) Frequency histograms for bound mass fraction, $f$. Experiment
(grey) from \fref{granick_pmma}(b). Theory (empty), from
predicted distribution $P(f)\twid f^{-4/5}$ with $f_{\rm min} < f <
\omega$, where values for $f_{\rm min} = 0.09$ and $\omega = 0.47$
were derived from ref.  \cite{schneider:granick_bimodal}.}
\end{figure}

Motivated by the experiment of refs.
\cite{schneider:granick_bimodal,douglas:kinetics_pol_ads_jpc} on PMMA
adsorption, O'Shaughnessy and Vavylonis
\cite{ben:chemiphysi_letter,ben:chemiphysi_euro} attempted to
calculate layer structure at the detail of distributions of loop sizes
and contact fractions, $\Omega(s)$ and $P(f)$.  In this scaling
theory, early arriving chains formed a flattened surface monolayer
while late arrivers were visualized to adsorb irreversibly onto the
few available unconnected empty surface sites (see
\fref{late_stages}(a)).  Following ref.
\cite{ben:chemiphysi_letter}, if each chain-surface adhesion point
involves $\ncont$ attached monomers, then the surface density of free
``supersites'' (unoccupied surface patches large enough to accommodate
$\ncont$ monomers) is $\rhosuper \approx \Delta\Gammabound/\ncont$
where $\Delta\Gammabound \equiv \Gammaboundinf - \Gammabound$ is the
density of available surface sites and $\Gammaboundinf$ the asymptotic
density of surface-bound monomers.  As the surface approaches
saturation, the separation between supersites $\lsep \approx
\rhosuper^{-1/2}$ becomes large and the minimum loop size $s$ which
can come down just connects two free supersites, \ie $a s^{3/5} =
\lsep$ whence $s=(\ncont/a^2 \Delta\Gammabound)^{5/6}$.  Chains
adsorbing at this stage end up as trains of $\ncont$ monomers
separated by loops of order $s$ units (see \fref{late_stages}(a)).  The
bound fraction of those chains is $f = \partial
\Delta\Gammabound/\partial\Delta \Gamma \approx \ncont/s$ for large
$s$, where $\Delta\Gamma$ is the deviation from the asymptotic
coverage $\Gammainf$. Integrating this process up to saturation gives
                                                \begin{eq}{pepper}
a^2 \Delta\Gammabound = \ncont (a^2 \Delta\Gamma/6)^6\comma \gap
                P(f) = Af^{-4/5} \gap (f \ll 1)
\end{eq}
where $A$ is a constant of order unity.  \Eref{pepper} describes
a tenuously attached outer layer (small $f$ values) of late arriving
chains, dressing the flattened inner layer (see \fref{late_stages}(b)).  Two
notable features of $P(f)$ are: a peak due to the outer layer at small
$f$ and a peak from the early stages centered at $f=\omega$.  These
results capture some essential features of the experimental data for
PMMA as shown in \fref{gamma_pf}.  Interestingly, the predicted
scaling forms for the loop distribution and density profile, $\Omega
(s) \approx a^{-2} s^{-11/5}$ and $c(z) \twid z^{-4/3}$, are the same
as those predicted for equilibrium layers (see
equations \eqref{equilibrium} and \eqref{input}).  However, a fundamental
distinction is that the predicted distribution of bound fractions,
\fref{gamma_pf}(b), is very different to the equilibrium one
shown in \fref{peq}(b).  This reflects the fact that, unlike
equilibrium, all chains are no longer statistically identical; rather,
there are infinitely many families, each parametrised by an $f$ value.

\setcounter{footnote}{0}

\section{Chemisorption}

Chemisorption is characterized by very small values of the local
monomer-surface reaction rate $Q$.  Despite the importance of chemical
adsorption in many applications, few experimental and theoretical
works exist.  Irreversibility effects are almost always relevant since
chemical bond energies are virtually always much larger than $kT$.

\subsection{Theory}

\begin{figure}[tb]
\centering
\resizebox{0.7\columnwidth}{!}{%
  \includegraphics{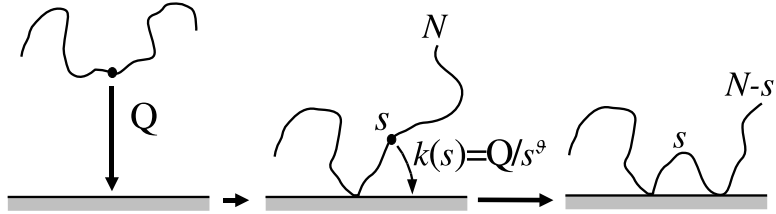}
}
\caption{Single chain chemisorption. A chain makes its first
attachment with the surface and begins its adsorption process.  The
rate of first monomer attachment is proportional to $Q$, namely the
reaction rate given the monomer touches the surface.  While for
physisorption subsequent monomer adsorption is diffusion-controlled,
for chemisorption the reaction rate $k(s)$ of the $s\nth$ monomer away
from the first attachment is proportional to the equilibrium
monomer-surface conditional contact probability. For $s \ll N$, where
$N$ is length of the tail, $k \approx Q/s^{\theta}$ where the exponent
$\theta$ reflects equilibrium polymer statistics near an interface
subject to the constraints imposed by earlier reactions. }
\label{single_chain_story}
\end{figure}

The important feature of chemisorption is that due to the smallness of
$Q$, chemical bond formation requires times which are typically longer
than the relaxation time of the attached chain.  Thus the reaction
rate $k(s)$ of a monomer $s$ units from a graft point is proportional
to the equilibrium probability of finding this monomer on the surface,
given the current constraints due to all chemical bonds formed at
earlier times, see \fref{single_chain_story}.  The kinetics of single
chain chemisorption were first studied by Shaffer et
al. \cite{shaffer:reactions_metal_polymer_jcp,%
shafferchakraborty:pmma_chemisorption_kinetics} who focused on PMMA
chemisorption onto Al.  The form of $k(s)$ corresponding to theta
solvent solutions was substituted into the master equation describing
the evolution of the chain's loop distribution.  Electronic structure
calculations \cite{shaffer:reactions_metal_polymer_jcp} indicated
enhanced reaction rates for monomers neighboring a graft point.  Due
to this local cooperative effect, the numerical solution of the master
equation led to a process of chain zipping onto the surface growing
outwards from the first attachments.

Ponomarev et al.  \cite{ponomarev:artem_durning} used the
bond-fluctuation model to simulate single chain physisorption and
chemisorption by varying monomer sticking energies and adsorption
barrier heights. For chemisorption, in the limit of small $Q$ they
found $\tauads \twid N^{0.8 \pm 0.2}$ with a magnitude greater than
the bulk chain relaxation time.

Scaling theory and asymptotic solutions of the chemisorbing chain
master equation were developed by O'Shaughnessy and Vavylonis
\cite{ben:chemiphysi_letter,ben:chemiphysi_euro}.  Three general modes
of adsorption were found, depending on value of the exponent
describing the decay of the reaction rate, $k(s) \twid s^{-\theta}$.
For $\theta > 2$ the total reaction rate, $\int_0^\infty ds\, k(s)$,
is dominated by small loop sizes and this was found to lead to a
zipping adsorption mode, the chain zipping down from the first
attachment point.  For $\theta < 1$, since the total reaction rate is
dominated by large $s$ large loops are more likely to form; this was
found to lead to a process of uniform chain collapse, very different
to zipping.  The case $1 < \theta < 2$, named ``accelerated zipping,''
was found to be intermediate between zipping and collapse; the pure
zipping process is effectively short-circuited by the occasional
adsorption of large loops which serve as nucleation points for further
zipping. For non-attractive surfaces exact results were obtained for
the exponent $\theta$ which can be expressed
\cite{ben:chemiphysi_euro} in terms of other known critical exponents
characterizing polymer networks
\cite{duplantier:networks,debelllookman:surface_exponents_review}.
For dilute solutions the relation $\theta = 1 + \nu$ was obtained,
where $\nu$ is Flory's exponent \cite{gennes:book}, giving
$\theta=8/5$ and $\theta=3/2$ for good and theta solvents,
respectively.  Thus the adsorption mode is accelarated zipping in both
cases, and the adsorption time was found to be $\tauads \approx Q^{-1}
N^{\theta -1}$.

O'Shaughnessy and Vavylonis
\cite{ben:chemiphysi_letter,ben:chemiphysi_euro} considered also the
kinetics of the build up of a many-chain chemisorbed layer from dilute
solution.  Arguing that the timescale for single chain adsorption was
much smaller than the time for adsorbing chains to interfere with one
another, they predicted a monolayer forms in the early stages.  This
is followed by a diffuse outer layer, attaching at the few available
empty surface sites, just as for physisorption, \fref{late_stages}.
The final layer structure was thus of the same form, with two chain
populations as in \fref{late_stages}(b) and with the same loop and
contact fraction distributions. However, the bound versus total mass
relation differs for small coverage, $\Gammabound \twid \Gamma^{8/3}$
as in \fref{gamma_pf}(a). The kinetics of layer formation are also
very different: in physisorption diffusion-control was predicted to
apply at all times, $\Gamma\twid t^{1/2}$, while for chemisorption
$\Delta\Gamma\twid t^{-1/5}$ at long times.

\subsection{Experiment}

In contrast to physisorption, very few experiments have examined the
fundamentals of polymer chemisorption.  In principle however, since
timescales are intrinsically much longer, the kinetics might be easier
to probe.  Schlenoff et al.
\cite{schlenoff:functionalized_ps_on_gold} studied PS with functional
thiol groups along its backbone chemisorbing on gold from THF
solution.  Surface coverage was measured as a function of time using
the quartz crystal microbalance and the degree of polymer
functionalization, $q$, was varied.  Pure PS did not adsorb; for $q =
1\%$ and MW=127,000, the asymptotic surface coverage was higher and
was reached faster than when $q=5\%$, MW=304,000.  Lenk et
al. \cite{lenk:functionalized_pmma_on_gold} studied
thiol-functionalized PMMA chemisorption on gold.  Similarly to the
results of ref. \cite{schlenoff:functionalized_ps_on_gold} it was
found that less polymer adsorbed with increasing $q$.  In every case
however the final adsorbed amount was higher than the amount of
unfunctionalized PMMA physisorbed onto gold.  The final layer was
found to contain unreacted thiols.  Tsao et
al. \cite{tsao:functionalized_pdms_on_gold} studied
thiol-functionalized PDMS adsorption onto gold from dichloromethane.
Similarly to refs.
\cite{schlenoff:functionalized_ps_on_gold,lenk:functionalized_pmma_on_gold}
less polymer adsorbed with increasing $q$.  However unlike the PMMA
system \cite{lenk:functionalized_pmma_on_gold} the final layer did not
contain unreacted thiols except for the case $q=1$.  These thiol
experiments are difficult to interpret at present since thiols not
only chemisorb via their S group, but are also very powerful
physisorbers onto gold \cite{lavrich:thiols_on_au}.  In reality,
therefore, a complex mixture of simultaneous physisorption and
chemisorption is involved.  It has also been suggested that they may
self-associate in solution \cite{schlenoff:functionalized_ps_on_gold}.

Cosgrove et al. \cite{cosgrove:chemisorption_ftir} studied the
simultaneous physisorption and chemisorption of poly(methylsiloxane)
on Al by Fourier-transform infrared specroscopy measuring total
adsorbed mass as a function of polymer concentration in the bulk.
Higher final coverages were found with increasing bulk concentration
for a fixed adsorption time of one week.  Physisorption processes
rapidly formed a layer on the surface in less than an hour while
chemisortion processes continued for days.  For this system it was
found that less mass was adsorbed when, prior to the experiment, the
surface was initially exposed to a solution of the same polymer at
lower concentrations.  Qualitative agreement with this experiment was
found in a lattice monte-carlo model developed in refs.
\cite{cosgrove:phys_and_chem_adsorption_langmuir,kingcosgrove:ads_pol_dynamical_monte_carlo}.
The simultaneous physisorption and chemisorption of poly(hydrogen
methylsiloxane) on quartz and porous silica was studied by neutron
reflectometry and small angle neutron scattering in ref.
\cite{cosgrove:chemisorption_neutron_scattering}.  Adsorption was
allowed to proceed for approximately one day at different
temperatures.  It was found that at higher temperatures the final
layer had a more extended density profile and the total coverage was
higher than at lower temperatures.

\setcounter{footnote}{0}
\section{Concentration Effects: Irreversible Adsorption from Melts and
Semi-Dilute Solutions}

In this section we review irreversible layer formation from polymer
melts and solutions at finite concentration.  This is a common
situation arising in many applications involving reinforcement of
polymer-solid or polymer-polymer interfaces
\cite{kraus:book_reinforcement_elastomers,wu:polymer_iface_adhesion_book,%
edwards:review_filler_reinforcement,creton:kramer:iface_fracture_review}
where, for example, surfaces may be functionalized by various types of
coupling agents
\cite{edwards:review_filler_reinforcement,cleggcollyer:book,%
pireaux:polymer_solid_ifaces_book}.  In general, interfacial
adsorption from melts and concentrated solutions may arise from either
physical or chemical interactions.  Clearly, the behaviour of chains
in the immediate vicinity of a surface is relevant and such effects
have been probed in experiments studying non-equilibrium polymer
dynamics in thin melt films in contact with surfaces
\cite{jones:thin_films_review_99,jones:chain_thin_film_nature_99,
forrest:thin_films_review_epje_02}.
Here we we do not review this large and developing reasearch field.

\subsection{Physisorption: The Guiselin Brush}

\subsubsection*{Theory.}
\begin{figure}[tb]
\centering
\resizebox{0.7\columnwidth}{!}{%
\includegraphics{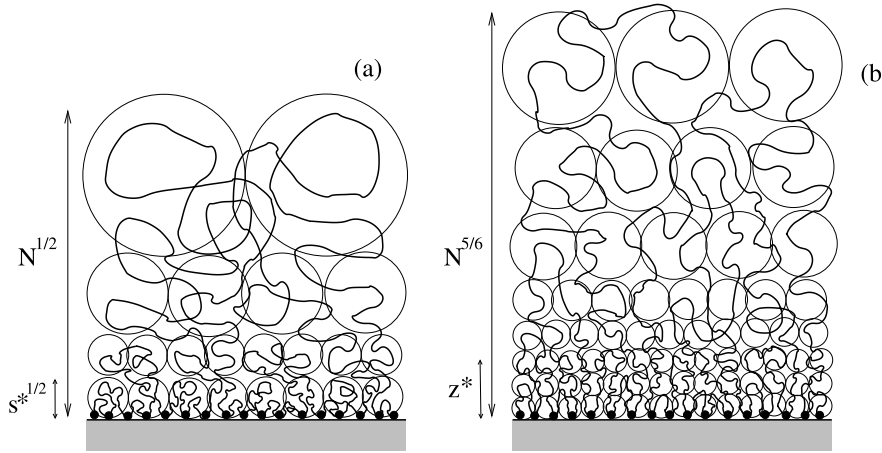} }
\caption{Irreversibly adsorbed layers from polymer
melts. Theoretically predicted structures for physisorption and
chemisorption are rather similar except at small scales. (a)
Hierachichal loop structure after adsorption from the melt.  For slow
chemisorption \cite{ben:chemiguiselin_letter} successive loop size
scales are frozen in as their critical density is reached.  Beyond
this their formation would force chain stetching and hence is strongly
suppressed.  After time $t$ the loops bigger than $\sstar \twid
1/(Qt)^2$ get frozen in.  For irreversible physisorption
\cite{guiselin:irrev_ads} by comparison all loop sizes freeze
immediately.  The end result is the same loop distribution but with
$\sstar$ essentially one unit.  (b) The experiment proposed by
Guiselin \cite{guiselin:irrev_ads} entails swelling the adsorbed
structure shown in (a) with solvent and washing away unattached melt
chains.  The blob structure of the swollen layer in good solvent is
shown.  It consists of two regions separated at $\zstar \twid
{\sstar}^{3/5}$ corresponding to loop size $\sstar$
\cite{ben:chemiguiselin_letter,guiselin:irrev_ads}, each region having
a different density profile.  For irrversible physisorption from a
melt there is no inner region. }
\label{blobs-swell}      
\end{figure}

Guiselin \cite{guiselin:irrev_ads} proposed the following experimental
procedure to study irreversible polymer adsorption from melts.  He
considered the situation where a melt is exposed to a surface so
attractive to the polymer chains that they adsorb instantaneously and
irreversibly.  This can be realized in practice for very strong
physisorption (\eg strong hydrogen bonding).  Assuming that chain
configurations are equilibrated before the melt is brought in contact
with the surface, then adsorption results in the irreversible freezing
in of the equilibrium configurations of the chains which touch the
surface.  Given the ideal chain statistics in a polymer melt
\cite{gennes:book}, the resulting loop distribution was found to be
                                                \begin{eq}{frozen}
\Omega(s) \approx 1/s^{3/2} \comma \gap (s < N^{1/2}) \period
\end{eq}
Here $\Omega$ is proportional to the probability that a random walk
originating at the surface makes its first return to the surface after
$s$ steps.  Guiselin proposed measuring this distribution by washing
away unattached chains with a good solvent (see \fref{blobs-swell}).
This results in a swollen and stretched polydisperse brush with a
density profile \cite{guiselin:irrev_ads}
                                                \begin{eq}{swollen}
a^3 c(z) \approx (a/z)^{2/5} \comma \gap (z < h = aN^{5/6}) \comma
\end{eq}
where $h$ is layer height.  An important difference between Guiselin's
layer and equilibrium layers in dilute solutions (see
\eref{equilibrium}) is in the value of the total surface coverage,
$\Gamma = \int_a^\infty dz\, c(z)$.  For equilibrium layers, from
\eref{equilibrium} one obtains $\Gamma \approx a^{-2}$ for $\epsilon >
kT$.  For the Guiselin brush on the other hand, most monomers in the
melt within the coil radius $aN^{1/2}$ from the surface belong to
chains which became attached.  This leads to much higher surface
coverages, $\Gamma \approx N^{1/2} a^{-2}$.  If the above melt
experiment is repeated in a semidilute solution of polymer volume fraction
$\phi$, Guiselin predicted that a fraction of order unity of monomers
within the coil radius\cite{gennes:book} $R \approx a N^{1/2}
\phi^{-1/8}$ from the surface will become attached.  This was shown by
Marques and Joanny \cite{marquesjoanny:semidil_ads_fractal} to lead to
a surface coverage
                                                \begin{eq}{everything-sticks}
a^2 \Gamma \approx (R/a) \phi \approx N^{1/2} \phi^{7/8} \comma
                                                                \end{eq}
from which the melt case is recovered by setting $\phi = 1$ (see also
ref. \cite{johner:chain_statistics_adsorbed_pol}).  For distances
larger than the bulk correlation length of the semidilute solution, $z
> a \phi^{-3/4}$, the density profile after removal of unattached
chains was predicted to decay as $z^{-2/5}$ as in \eref{swollen}.

The structure of Guiselin's brush in contact with a polymer solution
has been studied theoretically by Aubouy and Rapha\"el
\cite{aubouyraphael:guiselin_brush_polsolution} and experimentally by
Auroy and Auvray \cite{auroyauvray:brush_polsolution_sans}.  Adhesion
properties have been reviewed by L\'eger et al.
\cite{leger:anchored_chains_adhesion_review}.

\subsubsection*{Numerical Simulations.}

The kinetics of irreversible physisorption from a semi-dilute solute
were studied by Monte Carlo simulations by Zajac and Chakrabarti
\cite{zajacchakrabarti:irrev_pol_ads_monte_carlo} assuming
instantaneous and irreversible adsorption of monomers upon contact
with the surface (values of $N$ employed were 50,100, and 200).  It
was found that for long times the surface coverage reached values
consistent with an approximate law $\Gamma \twid N^{0.38}
\phi^{0.38}$.  The density profile of the adsorbed chains within the
semidilute solution (\ie without removing unattached chains) appeared
to decay more slowly than $z^{-2/5}$.  The first moment of the profile
was found to increase as $\phi^{0.20 \pm 0.03} N^{0.81 \pm 0.05}$.  It
was found that the layer consisted of two regions: a loop-dominated
inner region and a tail-dominated outer region.

Jia and Lai \cite{jialai:irrev_pol_ads_monte_carlo} used the
bond-fluctuation Monte Carlo model to study instantaneous and
irreversible polymer adsortpion from semi-dilute solutions.  Contrary
to Zajac and Chakrabarti
\cite{zajacchakrabarti:irrev_pol_ads_monte_carlo} they found that
$\Gamma$ asymptotes its long time value with a stretched exponential
time-dependence.  Their results for layer height of the swollen layer
were consistent with Guiselin's prediction in \eref{swollen},
though the chain lengths which were used were too short to probe the
density profile ($N \le 50$).  They observed loop and tail regions
similarly to ref. \cite{zajacchakrabarti:irrev_pol_ads_monte_carlo}.

\subsubsection*{Experiment.}

Experimentally, Cohen-Addad
\cite{cohenaddad:pdms_melt_silica_coverage_1989} and Cohen-Addad and
Dujourdy \cite{cohenaddaddujourdy:pdms_melt_silica_kinetics_1998}
established that large surface coverages can be obtained in adsorption
from melts.  These studies involved hydrogen bonding of PDMS melts of
various molecular weights onto fumed silica particles.  The particles
were dispersed in the melt for varying periods of time.  After washing
away unattached chains with solvent, drying, and weighing, $\Gamma$
was extracted as a function of $N$.  For large enough times,
\eref{everything-sticks} was satisfied with $\phi = 1$.  However the
asymptotic surface coverage was obtained after a very long time
(weeks).  These slow kinetics were interpreted within a model in which
water molecules are initially adsorbed onto silica and prevent PDMS
monomers from adsorbing; the kinetics was thus controlled by the
diffusion of water molecules away from the surface
\cite{cohenaddadgennes:guiselin_brush_poison_elimination,%
cohenaddadmorel:pdms_melt_silica_kinetics_1995,%
cohenaddaddujourdy:pdms_melt_silica_kinetics_1998}.  Other models have
also been suggested \cite{levresse:pdms_melt_silica_mf_kinetics}.

Auvray et
al. \cite{auvray:irrev_ads_neutron,auvray:irrev_ads_concentrated_solution}
used small angle neutron scattering to study physisorption of PDMS
onto silica through strong hydrogen bonding.  Samples were prepared by
immersing porous silica in a PDMS melt or a semidilute solution of PDMS
in dichloromethane for a few days.  The samples were then washed with
dichloromethane and measured.  By varying both $N$ and the concentration
of the semidilute solution, their measurements of surface coverages were
in good agreement with \eref{everything-sticks}.  Measurements of
the density profile for adsorption from a melt showed layer heights
consistent with $h \twid N^{5/6}$ and evidence was found that the
density profile could indeed decay as a power law.

Durning et al. \cite{ben:pmma_adsorption} used neutron reflectivity to
study physisorption of PMMA from melts onto flat surfaces of
hydroxylated quartz via hydrogen bonding.  The melt was spin-cast onto
the quartz plates and annealed for different time periods at 165$^o$C.
Unbound chains were subsequently washed with benzene after cooling and
the sample was measured.  The total irreversibly adsorbed amount
$\Gamma$ was found to increase with annealing time, reaching a plateau
after a few days.  Since the process of spin-casting does not
immediately generate equilibrium chain configurations at the
interface, the dependence of $\Gamma$ on annealing time was intepreted
to be due to slow equilibration processes occuring at the interface
during annealing.  For the largest annealing times a scaling $\Gamma
\twid N^{0.47 \pm 0.05}$ was found, consistent with
\eref{everything-sticks}.  Their measurements of the density
profile of the swollen layer were consistent with a power law decay,
with an exponent whose value was about twice the predicted value of
\eref{swollen}, while measured layer heights were smaller, $h
\twid N^{0.72 \pm 0.03}$.

Marzolin et al. \cite{marzolin:eng_graft_and_guiseling_brush} also
used neutron reflectivity to study hydrogen-bonding of PDMS and
deuterated PDMS (dPDMS) onto silicon wafers.  Wafers were incubated in
a PDMS melt for 12-24 h at 110$^oC$ and unattached chains were rinsed
by solvent (deuterated toluene for PDMS and octane for dPDMS).
Measurements of density profiles for dPDMS showed good agreement with
Guiselin's prediction, \eref{swollen}, but PDMS layers appeared
to be less swollen.

The above experiments raise two important issues which deserve further
study: (1) Guiselin's thought experiment which envisions the
instantaneous freezing-in of equilibrium melt configurations in
practice may require annealing over large timescales
\cite{ben:pmma_adsorption} for many experimental realizations.  If the
Guiselin predictions are to apply, this timescale, which depends on
these unknown chain dynamics, must be large enough for any
non-equilibrium configurations generated at the interface during melt
deposition to have dissapeared.  (2) After swelling by solvent, one
expects desorption-readsorption events and slow chain movements.  This
will eventually transform \eref{swollen} into the equilibrium
profile of \eref{equilibrium}.

\subsection{Chemisorption: The Slow Guiselin Brush}

Since chemical bond formation requires many surface-monomer
collisions, chemisorption from melts can be thought of as a slow
version of Guiselin's case which involved immediate bond formation.
This process has been studied theoretically by O'Shaughnessy and
Vavylonis \cite{ben:chemiguiselin_letter}.  Single
chain chemisorption kinetics was analyzed in a similar way to that for
dilute solutions (section 5.1).  The chief difference is the reaction
exponent $\theta$ determining the surface reaction rate $k(s)$ of the
$s\nth$ monomer from a graft point (see \fref{single_chain_story}).
The exponent $\theta$ now represents ideal melt statistics due to
screening of excluded volume interactions \cite{gennes:book} leading
to a value $\theta=1/2$.  Values of $\theta$ less than unity lead to
{\em collapse} kinetics, during which each loop splits into two
daughter loops of approximately equal size, very different to zipping
mode.  Since the lifetime of a loop of length $s$ is $\taus \approx
1/\int_0^s ds' k(s') \approx 1/(Qs^{1/2})$, by time $t$ the maximum
loop size which could have survived is $\sstar = 1/(Qt)^2$.  Thus at
time $t$ a single chain consists roughly of $\twid N/\sstar$ loops of
size $\sstar$.

The single chain collapse kinetics can only be realized for very
dilute reactive chains in a bulk of identical but unreactive chains.
When all chains are functionalized, it was shown in ref.
\cite{ben:chemiguiselin_letter} that uniniterrupted
single chain collapse does not proceed to completion.  The reason is
that if all $1/N^{1/2}$ chains per surface site were able to collapse
independently, each would generate $N/\sstar$ loops of size $\sstar$
by time $t$, giving a net density $\approx N^{1/2}/\sstar$.  This
would exceed the critical level, $1/{\sstar}^{1/2}$, beyond which
adsorbed loops would have to be strecthed
\cite{gennes:grafted_pol_macro}.  In refs.
\cite{ben:chemiguiselin_letter} it was assumed that
loops are generated up to the critical density and no further.  This
leads to a layer which at any time consists of two parts: (i) an outer
layer of frozen large loops at the critical density, and (ii) an inner
layer of loops which have not yet collapsed.  If reactions are stopped
at $\tfinal$ and the layer is swollen with solvent, the predicted
layer height is $h = a N^{5/6}$ and the swollen layer has two regions
with different density profiles $c(z)$: $c(z) \twid {\zstar}^{-2/5}$ for
$z<\zstar$ and $c(z) \twid z^{-2/5}$ for $\zstar<z<h$ where $\zstar =
(Q\tfinal)^{-5/3}$.  Note that the outer layer has the same power law
profile and height as Guiselin's, \eref{swollen}, but the inner
part is different.

\setcounter{footnote}{0}
\section{Discussion}

The experimental studies of strong physisorbing systems from solution
in refs.\cite{schneider:granick_bimodal,douglas:kinetics_pol_ads_jpc}
have identified an important layer property, namely the distribution
of chain bound fractions, $P(f)$.  Features of this fundamental layer
characteristic may be more accessible to experiment than density
profiles.  In the appendix it was shown that for equilibrium layers in
the long chain limit this distribution has a sharp peak at a value of
$f$ of order unity.  Theoretically, then, the equilbrium layer
consists of tightly bound and statistically identical chains, with the
exception of a few more weakly attached chains (which however determine
layer height).  This general picture is supported by computer
simulations \cite{wangrajagopalan:pol_ads_dynamics_monte_carlo,
zajacchakrabarti:pol_ads_statics_dynamics_monte_carlo} which show that
in practice for finite length chains the peak of $P(f)$ may be rather
broad.  The experimental discovery of a qualitatively very different
and broad $P(f)$ for the strongly physisorbing PMMA-oxidized silicon
system \cite{schneider:granick_bimodal,douglas:kinetics_pol_ads_jpc}
has demonstrated that such non-equilibrium layers comprise many
classes of chain, each with its own $f$ value corresponding to its
time of arrival at the surface.  In particular, a large subpopulation
consists of weakly bound late arriving chains.  The existence of a
broad $P(f)$ is supported by theories addressing irreversible
adsorption
\cite{douglas:kinetics_pol_ads_jpc,ben:chemiphysi_euro,ben:chemiphysi_letter}.

A promising approach for future experiments would be to probe $P(f)$
and its dynamics for less sticky systems which exhibit measurable
aging effects, such as PS on oxidised silicon
\cite{frantzgranick:kinetics_adsorption_desorption_prl,
frantzgranick:ps__cyclohexane_exchange_macro,schneidergranick:kinetic_traps_exchange}.
It would be interesting to test what the initial $P(f)$ is in this
case and if it does tend to relax towards equilibrium.  During this
relaxation process, the late arriving, weakly bound chains will
progressively become more attached to the surface.  Since weakly bound
chains are presumably more easily displaced by bulk chains, it is
possible that it is the relaxation process of $P(f)$ which underlies
the slowing down of exchange rates with increasing aging in many of
the displacement experiments (see section 3.3).

Another promising future probe of processes in non-equilibrium layers
lies in measurements of the dynamics of the lateral size of individual
chains.  As the chains which build up the layer flatten down onto the
surface, theory has suggested that their initial size is of order the
bulk size, $R \twid N^{3/5}$.  In equilibrium layers however,
theoretical arguments \cite{semenovjoanny:kinetics_adsorption_rouse}
support a picture in which most chains have a latteral size $R \twid
N^{1/2}$ with the exception of a small fraction of weakly attached
chains extending far into the bulk.  This leads one to expect a
process of chain shrinkage during layer aging by a factor which may
typically be of order unity.  This change could be measurable with
smart chain labeling.

Recently, Zhao and Granick \cite{zhaogranick:layer_diffusion_jacs_04}
measured chain diffusion coeffients, $D$, as a function of surface
coverage, $\Gamma$.  They employed PEO chains with MW=10800 adsorbed
onto rather weakly attractive surfaces for which they estimated
$\epsilon \approx 0.5 kT$.  They found that $D$ was enhanced when the
surface coverage increased towards what seemed to be the chain overlap
threshold.  But for layers whose coverage was far above this limit,
they observed a large reduction in $D$.  An interesting question
motivated by this experiment is understanding how the curve
$D(\Gamma)$ is modified with increasing surface stickiness.

The differing structures of the equilibrium and non-equilibrium layers
should have important consequences for layer mechanical properties and
interactions between layers.  In the picture of refs.
\cite{ben:chemiphysi_letter,ben:chemiphysi_euro} the scaling of the
density profile is the same for fully formed equilibrium and
irreversibly formed layers.  Now the interactions between two
polymer-covered surfaces are believed to depend primarily on the
density profiles
\cite{gennes:polads_force_layers_macro_82,rossipincus:starved_pol_layers_macro,aubouy:scaling_flat_layers}.
Thus, when two such surfaces are brought into contact using a surface
force apparatus \cite{raviv:polymer_mat}, for example, this suggests force profiles
should be qualitatively similar regardless of whether the polymer
layers formed reversibly or irreversibly.  Since the equilibrium layer
chains rearrange themselves relatively rapidly, differences may
manifest themselves at long times or very high forces when significant
chain rearrangements can occur
\cite{raviv:polymer_mat,johnersemenov:comment_polymer_mat,raviv:reply_comments_polymer_mat},
leading to chain bridging and other complex effects.  On the other
hand adhesion properties may be very different in the two cases.
Suppose a polymer layer is brought into contact with an elastomer or a
polymer melt which is subsequently cooled
\cite{leger:anchored_chains_adhesion_review}.  Even though equilibrium
and irreversible layers have similar loop distributions, the statistics
of surface-chain attachments are very different (see sections 2.2.2
and 4.2). Consider a loop size $s$ large enough to efficiently couple
the surface to the bulk phase.  For an equilibrium layer, this loop
belongs to a chain having order $N$ contacts with the surface; for an
irreversibly formed layer the number of contacts is of order $N/s$,
since the loop distribution for a given chain is roughly monodisperse.
This suggests the equilibrium layer may produce a significantly
stronger interface, though the details are complex and depend on the
mode of fracture \cite{creton:kramer:iface_fracture_review}.

Finally, the process of polymer chemisorption remains largely
unexplored experimentally and many possibilities for future work
exist.  While many chemisorbing systems may in practice involve some
degree of simultaneous physisorption, there are also certain
simplifications since the chemisorption process is very slow and often
very close to the clean limit of complete irreversibility.  Future
research on the structure of chemisorbed layers promises to yield
important insights into the many chemisorption-based applications
involving reinforcement of polymer interfaces and adhesion.


\ack

This work was supported by the National Science Foundation under grant
no. DMR-9816374.


{\appendix
\section{Calculation of Equilibrium Distribution of Bound Fractions, $\Peq(f)$}

In this Appendix we calculate $\Peq(f)$ for a chain of length $N$ in
an equilibrium layer based on the statistics of loops and tails.  We
will make the approximation that the partition function $Z(n|N)$ of a
chain of length $N$ making $n$ contacts with the surface (\ie having 2
tails and $n-1$ loops) is the product of $n+1$ ``one step'' loop and
tail weighting factors (see end of appendix for a self-consistency
argument).

{\bf Definition of partition functions.}
Denoting $\Wt(s|N)$ and $\Wl(s|N)$ the weighting factor for a tail and
loop of size $s$, respectively, we have:
                                                \begin{eq}{zzz}
\fl
Z(n|N)  =
\int_0^N dt_1 \int_0^N  dt_2 \prod_{i=1}^{n-1}  \square{ \int_0^N ds_i}
\delta\paren{N-t_1-t_2-\sum_{i=1}^{n-1} s_i}
\Wt(t_1|N) \Wt(t_2|N)  \prod_{i=1}^{n-1} \Wl(s_i|N) \period
\drop
\end{eq}
The authors of refs.
\cite{semenovjoanny:loops_tails_europhys,semenovjoanny:kinetics_adsorption_rouse}
evaluated the scaling forms of the partition functions $\Zl(s|N)$ and
$\Zt(s|N)$ of loops and tails in the layer, respectively, which
generate the self-similar equilibrium density profile.  Thus the
properties of $\Wt$ and $\Wl$ in \eref{zzz} must be such that
                                                \begin{eqarray}{sj}
\Zt(s|N) &=& \Wt(s|N) \sum_{n=1}^{\infty} \Zbar (n|N-s) \comma
\drop
\Zl(s|N) &=& \Wl(s|N) \sum_{n=1}^{\infty} \sum_{m=1}^{\infty}
\int_0^{N-s} ds_1 \Zbar(n|s_1) \Zbar(m|N-s-s_1) \comma
\end{eqarray} 
where $\Zbar$, defined similarly to \eref{zzz}, is the partition
function for a chain with with just one tail (its other end being in
contact with the surface).  We now make the ansatz that $\Wl$ and
$\Wt$ have the same power law scaling as $\Zl$ and $\Zt$,
respectively, and we will later verify self-consistently that
\eref{sj} is indeed satisfied in the limit $N \gt \infty$.  Using the
results of refs.
\cite{semenovjoanny:loops_tails_europhys,semenovjoanny:kinetics_adsorption_rouse}
we thus consider $\Wl(s|N) = \mu^s \Al \eta_l(s|N)$ and $\Wt(s|N) =
\mu^s \At \eta_t(s|N)$.  Here $\Al$ and $\mu$ are constants of order
unity, $\At \approx N^{(\gamma-\nu)/4\nu}$, and
                                                \begin{eqarray}{partition}
\eta_l(s|N) 
\approx \casesbracketsshortii
                     {s^{-1-2\nu}}{s\ll\sstar}
                     {{\sstar}^{-1-2\nu}(\sstar/s)^{1+5\nu+\gamma}}{s \gg
                     \sstar}             
\drop
\eta_t(s|N)
\approx \casesbracketsshortii
                     {{\sstar}^{-1}(\sstar/s)^{1-(\gamma-\nu)/2}}{s \ll \sstar}
                     {{\sstar}^{-1}(\sstar/s)^{1+2\nu}}{s \gg \sstar}
\gap \sstar \equiv N^{1/(2\nu)} \period 
\end{eqarray}
The functions $\eta_l$ and $\eta_t$ are normalised to unity,
$\gamma\approx 1.16$ is the susceptibility exponent
\cite{gennes:book}, $s$ is continuous, and a small-$s$ cutoff at $s$
of order unity is understood. 

{\bf Calculation of $Z(n|N)$.}
We introduce a new variable $M$ in \eref{zzz} which allows
us to simplify it by Laplace transforming $M\gt E$ as follows:
                                                \begin{eqarray}{zzz2}
\fl
Z(n|N) &=& \mu^N \At^2 \Al^{n-1} \Xi_N(n|N)  \comma
\drop
\fl
\Xi_M(n|N)  &\equiv&
\int_0^M dt_1 \int_0^M  dt_2 \prod_{i=1}^{n-1}  \square{ \int_0^M ds_i}
\delta\paren{M-t_1-t_2-\sum_{i=1}^{n-1} s_i}
\eta_t(t_1|N) \eta_t(t_2|N)  \prod_{i=1}^{n-1} \eta_l(s_i|N)  \period
\end{eqarray}
Since $\Xi_M$ is a convolution over loop and tail lengths, one has
                                                \begin{eqarray}{profilin}
\Xi_E(n|N) &=& [\eta_t(E|N)]^2[\eta_l(E|N)]^{n-1} 
\drop
&=& e^{(n-1)\ln\eta_l(E|N)+2\ln\eta_t(E|N)}
\drop
&\approx& e^{(n-1)[\eta_l(E|N)-1]+2\ln\eta_t(E|N)}
\gap (E \ll 1)
\comma
\end{eqarray}
where we used the fact that for $E \ll 1$, one has $\eta_l(E|N) \gt 1$.
Because the quantity $\eta_l(E|N)-1$ appearing in \eref{profilin} is
small-scale dependent (due to the fact that the exponent 1+2$\nu$ in
\eref{partition} is larger than 2), it is more natural to evaluate
instead the Laplace transform of the distribution $\Xi_Y'(n|N)$, where
                                                \begin{eq}{y}
\fl
Y \equiv M - (n-1) \, \sbarl - 2 \sbart \comma \gap
\sbarl \equiv
\int_0^\infty  ds\, s \, \eta_l(s|N) \comma \gap
\sbart \equiv
\int_0^\infty  ds\, s \, \eta_t(s|N) \approx \sstar \period
\end{eq}
Note that the loop size $\sbarl$ is of order unity and depends on the
cutoff of $\eta_l$ in \eref{partition}, while $\sbart \twid
N^{1/(2\nu)}$.  As will become clear below (\eref{invert}) in the
limit $M\gt\infty$ the probability of negative $Y$ values becomes
zero.  For $M \gg 1$ one can neglect negative $Y$ and after using
\eref{profilin} one has
                                                \begin{eq}{transform}
\fl
\Xi_E'(n|N) = \Xi_E(n|N) e^{ E\, (n-1) \sbarl + 2 E \sbart } =
e^{(n-1)[\eta_l(E|N)-1+E \sbarl]+2[\ln\eta_t(E|N)+E\sbart]}
\gap (E \ll 1) \comma
\end{eq}
where $E$ is now conjugate to $Y$.  Laplace transforming
\eref{partition} one obtains the following expressions for the
quantities appearing in \eref{transform}:
                                                \begin{eqarray}{zetapart}
1-\eta_l(E|N)-E \sbarl \approx 
\casesbracketsshortii{E^{2\nu}}{1\ll E^{-1}\ll\sstar}
                {{\sstar}^{-2\nu}(E \sstar)^{2}} {E^{-1}\gg \sstar}
\drop
-\ln\eta_t(E|N) -E\sbart \approx
\casesbracketsshortii{-(E\sstar)}{1\ll E^{-1}\ll\sstar}
                {(E\sstar)^{2\nu}} {E^{-1}\gg \sstar}
\end{eqarray}
Using \eref{zetapart} in \eref{transform} one finds that the
loop and tail contributions dominate for large and small $E$,
respectively:
                                                \begin{eq}{earth}
\Xi_E'(n|N) \approx 
\casesbracketsshortii
{ e^{-B(n-1)E^{2\nu}}}{1\ll E^{-1} \ll \sstar}
{e^{-C(E\sstar)^{2\nu}}}{E^{-1} \gg \sstar}
\end{eq}
where $B$ and $C$ are positive constants of order unity.  In deriving
\eref{earth} we have considered $f=n/N$ of order unity since,
self-consistently, the probability of very small $f$ values is very
small (see \eref{peq}).  Laplace inverting one has
                                                \begin{eq}{invert}
\Xi_Y' (n|N) \approx 
{N \over Y^{1+2\nu}}\ \zeta\paren{{Y\over n^{1/(2\nu)}}} \comma \gap
\zeta(x) \gt \casesbracketsshortii{1}{x \gg 1} 
                                 {0}{x \ll 1}\comma
\end{eq}
where $\zeta(x)$ is a cut-off function.  This distribution has a peak
at $Y \approx n^{1/(2\nu)}$, a long tail for large $Y$, and is
exponentially small for $Y\ll n^{1/(2\nu)}$.  One thus obtains
$Z(n|N)$ by using $Y\gt N - (n-1)\sbarl - 2\sbart$ in \eref{invert}
and substituting into \eref{zzz2}:
                                                \begin{eq}{green_sparrow}
Z(n|N) \approx \mu^N \At^2 \Al^{n-1} 
{N \over [N - (n-1)\sbarl - 2\sbart]^{1+2\nu}}\
 \zeta\paren{{N - (n-1)\sbarl - 2\sbart\over n^{1/(2\nu)}}} \period
                                                                \end{eq}
This expression will immediately yield $\Peq(f)$ (see below).

{\bf Self-consistency check and the value of $\Al$.} We now show that
our choice of weighting factors in \eref{zzz} leads to loop and tail
partition functions in \eref{sj} which are consistent with the
findings of refs.
\cite{semenovjoanny:loops_tails_europhys,semenovjoanny:kinetics_adsorption_rouse}
provided the normalisation constant $\Al$ in the 1-loop distribution
function obeys certain constraints.

Consider first $\Zt$ in \eref{sj}. Repeating the above calculations of
equations \eqref{zzz}-\eqref{invert} for the case of a chain with one
tail, one can show that $\Zbar(n|N)= \mu^N \At \Al^{n-1}
\Xibar_N(n|N)$, where $\Xibar_N$ has the same scaling structure as
$\Xi_N$.  Using the definition of $\Wt(s|N)$ one thus has from
\eref{sj}:
                                                \begin{eq}{tail}
\Zt(s|N) = 
\mu^N \At^2 \eta_t(s|N)
{\sum_{n=1}^{\infty} \Xibar_{N-s} (n|N-s) e^{(n-1)\ln \Al}} \period
\end{eq}
Now suppose $e^{N\ln \Al} \gg 1$, in which case the exponential
favours large values of $n$.  Since $\Xibar_{N-s}$ has a sharp cut-off
for $n$ values of order $(N-s)/\sbarl$, thus the sum in \eref{tail} is
then dominated by $n$ values of this order.  This would imply
$\Zt(s|N)\twid e^{-(s\ln \Al)/\sbarl }$ is exponentially suppressed for
larger $s$, inconsistent with the tail partition function derived in
\cite{semenovjoanny:loops_tails_europhys,semenovjoanny:kinetics_adsorption_rouse}.
The latter has a power law dependence on $s$ (note that $\Zt$ in
\eref{sj} involves a sum over configurations of the whole chain and
unlike $\Wt$ it involves a factor $\mu^N$ instead of $\mu^s$).  Thus,
self-consistency requires that the value of $\Al$ is such that
$e^{N\ln \Al}$ is non-diverging as $N \gt \infty$.

The case $e^{N\ln \Al} \ll 1$ is also inconsistent; then the chain
partition function, $Z(n|N)$, would be largest for $n$ values much
less than $N$ (see \eref{green_sparrow}).  This would imply that the
mean bound fraction is much less than unity, inconsistent with the
equilibrium layer for which a fraction of order unity of the adsorbed
mass contacts the surface. 

Overall, we conclude $e^{N\ln \Al}$ is of order unity.  In this case
the sum in \eref{tail} is to leading order independent of $s$ and thus
$\Zt(s|N)$ has the same scaling as $\eta_t(s|N)$; this is consistent
with the results of refs.
\cite{semenovjoanny:loops_tails_europhys,semenovjoanny:kinetics_adsorption_rouse}.
Repeating the above arguments for the case of loops, and considering
$e^{N\ln \Al}$ of order unity, one finds $\Zl(s|N)$ has the same
scaling with respect to $s$ as $\eta_l(s|N)$.  This is also consistent
with refs.
\cite{semenovjoanny:loops_tails_europhys,semenovjoanny:kinetics_adsorption_rouse}.

{\bf The distribution of bound fractions.}  The probability
$P(n|N)$ to have $n$ loops is obtained from $P(n|N) =
Z(n|N)/\sum_{n=1}^{N} Z(n|N)$.  Since the bound fraction is $f = n/N$,
then using $\Peq(f) df = P(n|N) dn$ and $\nu\approx 3/5$ one has
                                                \begin{eq}{almost}
\Peq(f) \approx X e^{f N\ln\Al} {N^{-1/5} \over (\fmax - f)^{11/5}} 
\ \rho \square{(\fmax-f) N^{1/6}}  \comma
\end{eq}
where $X$ is a normalisation constant, $\fmax \approx 1/\sbarl$, and
$\rho$ is defined in \eref{peq}. Since $e^{N\ln \Al}$ is of order
unity, from \eref{almost} one recovers \eref{peq} of the main text.

{\bf Self-consistency of \eref{zzz}}
The assumption of independent loop statistics is consistent with the
hypothesis of ideal lateral blob statistics at every scale in the
adsorbed layer \cite{semenovjoanny:kinetics_adsorption_rouse}.  The
approximation is also self-consistent in the sense that the
probability that a chain makes two overlapping loops of the same size
$s$ at the same spot on the surface is small.  The number of
successive loops $n$ needed to make a loop of size of order $s$ or
larger is $n \approx \int_s^N ds' \Omega(s') \approx s^{6/5}$.  During
these steps the chain would have performed a random walk on the
surface of size $n^{1/2} \approx s^{3/5}$.  Thus the origin of the
next loop of size $s$ is just outside the region occupied by the
previous loop of the same size.

}




\end{document}